\documentclass[referee]{raa}            

\usepackage{graphicx,times}             
\usepackage{natbib}
\usepackage{amssymb,amsmath}
\bibpunct{(}{)}{;}{a}{}{,}

\usepackage[pagebackref=true]{hyperref}

\begin{document}

  \title{Magnetic Activity Cycles and Rotation in Planet-Hosting and Non-Hosting Solar-Type Stars \thanks{Based on data obtained from the ESO Science Archive Facility}}

   \volnopage{Vol.0 (20xx) No.0, 000--000}      
   \setcounter{page}{1}          

   \author{Amina Boulkaboul
      \inst{1}
 \and Alessandro Sozzetti
      \inst{2}
 \and Caroline Soubiran
      \inst{3}
   \and Yassine Damerdji
      \inst{1,4}
        }

   \institute{CRAAG - Centre de Recherche en Astronomie, Astrophysique et Géophysique, Route de  l’Observatoire, Bp 63 Bouzareah, DZ-16340, Algiers, Algeria; {\it amina.boulkaboul@craag.edu.dz}\\
         \and INAF – Osservatorio Astrofisico di Torino, Via Osservatorio 20, I-10025 Pino Torinese, Italy\\ 
         \and Laboratoire d’Astrophysique de Bordeaux, Univ. Bordeaux, CNRS, B18N, allée Geoffroy Saint-Hilaire, 33615 Pessac, France \\
        \and
             Space sciences, Technologies and Astrophysics Research (STAR) Institute, Universit\'e de Li\`ege, Quartier Agora, All\'ee du 6 Ao\^ut 19c, B\^at. B5c, B4000-Li\`ege, Belgium\\
\vs\no
   {\small Received 20xx month day; accepted 20xx month day}}

\abstract{We analyze periodicities in radial velocity (RV) measurements and magnetic activity indicators (S-index and BIS) for 767 Gaia RV standard stars to distinguish between stellar activity and planetary signals. Significant RV periods were detected in only 359 of these stars. Rotation and magnetic cycle periods are identified through iterative periodogram analysis. Among stars with confirmed planets, 28.2\% exhibit RV signals that coincide with activity indicators, compared to $21.3\%$ among stars without planets; however, statistical tests show this difference is not statistically significant. Several RV signals previously attributed to planets—such as those in HIP7240, HIP28460, and HIP48331—are instead likely caused by stellar activity, emphasising the importance of using multiple diagnostics to assess RV variability. We report rotation periods in 125 stars, including 30 new estimates, and detect magnetic activity cycles in 127 stars, 95 of which are new. We further investigate the relationship between rotation and magnetic cycle periods in the context of stellar dynamo theory. The full sample reveals a continuous distribution in the $P_{\rm rot}/P_{\rm cyc}$–Rossby number diagram, lacking the classical division into active and inactive branches and instead showing a negative slope, in contrast to some earlier studies. Interestingly, stars with planetary companions exhibit a steeper trend (slope of $-1.049 \pm 0.078$) compared to non-hosts ($-0.654 \pm 0.056$), suggesting that the presence of planets may subtly influence the host star's magnetic behaviour.
}

   \authorrunning{A. Boulkaboul et al. }            
   \titlerunning{Magnetic Activity and Rotation in Late-Type Stars}  

   \maketitle

%
%
\section{Introduction}           
\label{sect:intro}

The discovery of the Sun’s 11-year activity cycle by \cite{schwabe1844sonnenbeobachtungen} was later extended to other late-type stars through systematic monitoring of chromospheric activity indicators such as \ion{Ca}{ii} H and K, \ion{Mg}{ii}, H$\alpha$, and the \ion{Ca}{ii} infrared triplet. A landmark in this field was the Mount Wilson survey \citep{1968ApJ...153..221W,1991ApJS...76..383D,1995ApJ...438..269B}, which tracked chromospheric activity in over a thousand cool stars for more than four decades. Early results by \cite{1978ApJ...226..379W} revealed that magnetic cycles, like the solar one, are common among stars with convective envelopes \citep{1985ARA&A..23..379B,1990ApJ...353..524R}, with a rich diversity including solar-like cyclic, highly variable non-cyclic, and flat activity \citep{boro2018chromospheric}.

Stellar magnetic cycles arise from dynamo processes generated by the interplay between rotation and convection \citep{1972ApJ...171..565S,noyes1984rotation}. Observationally, these processes manifest as surface features—such as starspots and plages—and long-term variability in chromospheric emission. To better understand stellar dynamo generation, the ratio of the stellar rotation period to the magnetic cycle period ($P_{\mathrm{rot}}/P_{\mathrm{mag}}$), as a function of the Rossby number (Ro = $P_{\mathrm{rot}}/\tau_c$, where $\tau_c$ is the convective turnover time) has been extensively studied \citep{noyes1984rotation,saar1999time,olah2009multiple}. For instance,\cite{noyes1984rotation} found that, for 13 main-sequence slow rotators from the Mount Wilson program, the magnetic cycle period scales with rotation period as $P_{\rm{mag}}\propto P_{\rm{rot}}^{1.25}$. This relation appears to vary with the star's activity level. In particular, \cite{1992ASPC...27..150S} identified two distinct "activity branches"—active and inactive—based on the $P_{\mathrm{rot}}/P_{\mathrm{mag}}$ ratio. Later, \cite{2007ApJ...657..486B} proposed that these two branches reflect the operation of different types of dynamos: cycles on the inactive branch are likely generated in deeper layers of the convection zone, while those on the active branch originate in shallower layers. 
While \cite{brandenburg2017evolution} found no clear correlation between the period ratio and the fractional depth of the convection zone, \cite{mittag2023revisiting} reinforced the findings of \cite{2007ApJ...657..486B}, confirming the bifurcation into activity branches. Although the existence of these branches has been questioned by \cite{boro2018chromospheric}, \cite{mittag2023revisiting} showed that this separation depends on the $B-V$ colour index, implying a relationship with both effective temperature and the star's position along the main sequence.

Intriguingly, the Sun lies between the two branches \citep{2007ApJ...657..486B,metcalfe2016stellar}, further highlighting the complexity of stellar dynamos. \cite{metcalfe2016stellar} suggested that the Sun may be in a transitional evolutionary state, with its 11-year cycle arising from a transitional dynamo. However, \cite{mittag2023revisiting} argued that the Sun’s two principal cycles are consistent with the two activity branches, implying that it is not a special or exceptional case. Some stars even display multiple magnetic cycles that span both branches \citep{berdyugina1998permanent,berdyugina2005spot}, suggesting evolving or coexisting dynamo modes.

Despite theoretical advances, the origin of multiple magnetic cycles, the physical basis of the activity branch separation, and the Sun’s peculiar position continue to challenge models. Observations have shown that while magnetic cycle periods generally decrease with faster stellar rotation \citep{noyes1984rotation,mascareno2016magnetic}, this trend can break down in both very active and very inactive stars \citep{giampapa2006survey}. These deviations may reflect changes in differential rotation regimes, such as a transition from solar-like to antisolar behaviour \citep{brandenburg2018enhanced,karak2020stellar}. Resolving these issues requires comprehensive long-term data across a range of stellar types.

Beyond their intrinsic interest, stellar magnetic cycles pose a significant challenge to exoplanet detection. With modern RV spectrographs now achieving cm/s precision \citep{pepe2011harps,pepe2013espresso}, activity-induced RV variability, on both rotational and magnetic timescales, has become a dominant noise source, often masking or mimicking planetary signals \citep{saar1997activity, dumusque2011harps}. Disentangling these signals from true Keplerian motion requires detailed knowledge of stellar variability and predictive models of its behaviour \citep{mascareno2016magnetic}. Meanwhile, missions such as Kepler \citep{borucki2010kepler} and TESS \citep{ricker2015transiting} have yielded detailed measurements of planetary systems—including planetary radii, masses, and orbits—as well as the properties of their host stars. These data offer new opportunities to investigate whether stars hosting planets differ from those without, particularly in terms of rotation and angular momentum.

An open question in this context is whether the presence of planets, especially massive ones, affects stellar rotation or magnetic activity. Several studies have explored whether star-planet interactions, occurring during the pre-main sequence phase or via tidal and magnetic effects, could slow stellar rotation or influence dynamo behaviour \citep{bolmont2019effect}. The observational evidence is mixed: some works find no significant effect \citep{ceillier2016rotation}, while others suggest that planet-hosting stars rotate more slowly \citep{gonzalez2015parent,sibony2022rotation}. Related studies point to broader angular momentum trends tied to stellar and planetary mass \citep{gurumath2018angular}, but a consistent physical interpretation is still lacking.

In this paper, we present estimations of stellar rotation periods and magnetic cycle periods for a sample of Gaia RV standard stars, a set of well-observed, bright, slowly rotating stars with stable RVs. By analysing these stars, both with and without known planetary companions, we aim to characterise their dynamo properties and investigate their placement on the active/inactive branches. Our goal is to test whether planet-hosting stars exhibit systematically different rotation-activity behaviours.

The structure of this paper is as follows. Section \ref{targ} presents an overview of the target stars. In Section \ref{method}, we describe the methodology used for RV measurements and chromospheric activity analysis. Section \ref{results} discusses the RV periodicities observed in magnetic activity indicators while Section \ref{sect:discuss} examines the activity branches of stars with and without planetary companions. Finally, Section \ref{conc} summarises our main conclusions based on these results.

\section{Target stars}
\label{targ}

The sample analysed in this study consists of Gaia RV standard star candidates from \cite{soubiran2018gaia}, specifically those flagged with CAL1 quality. These stars served as wavelength calibrators for Gaia Data Releases 2 and 3 \citep{brown2018gaia, vallenari2023gaia} and were selected based on their long-term RV stability, spectral suitability, and minimal variability—defined as having an RV standard deviation below $100\, ms^{-1}$ \citep{soubiran2013catalogue,sartoretti2018gaia}—to ensure precise RV measurements throughout the Gaia mission.

For our analysis, we only include stars that were observed with the HARPS (High Accuracy Radial velocity Planet Searcher) spectrograph \citep{mayor2003dall}, as its high resolution is essential for our measurements. The HARPS archive spectra used in this work are publicly available from the ESO database\footnote{\url{http://archive.eso.org/wdb/wdb/adp/phase3_spectral/form}}. HARPS provides a resolving power of $R = \lambda/\Delta\lambda =115 000$, covering the wavelength range $3780-6910 $\AA.
\begin{figure}
\begin{center}
\begin{subfigure}{.45\textwidth}
\includegraphics[width=\linewidth]{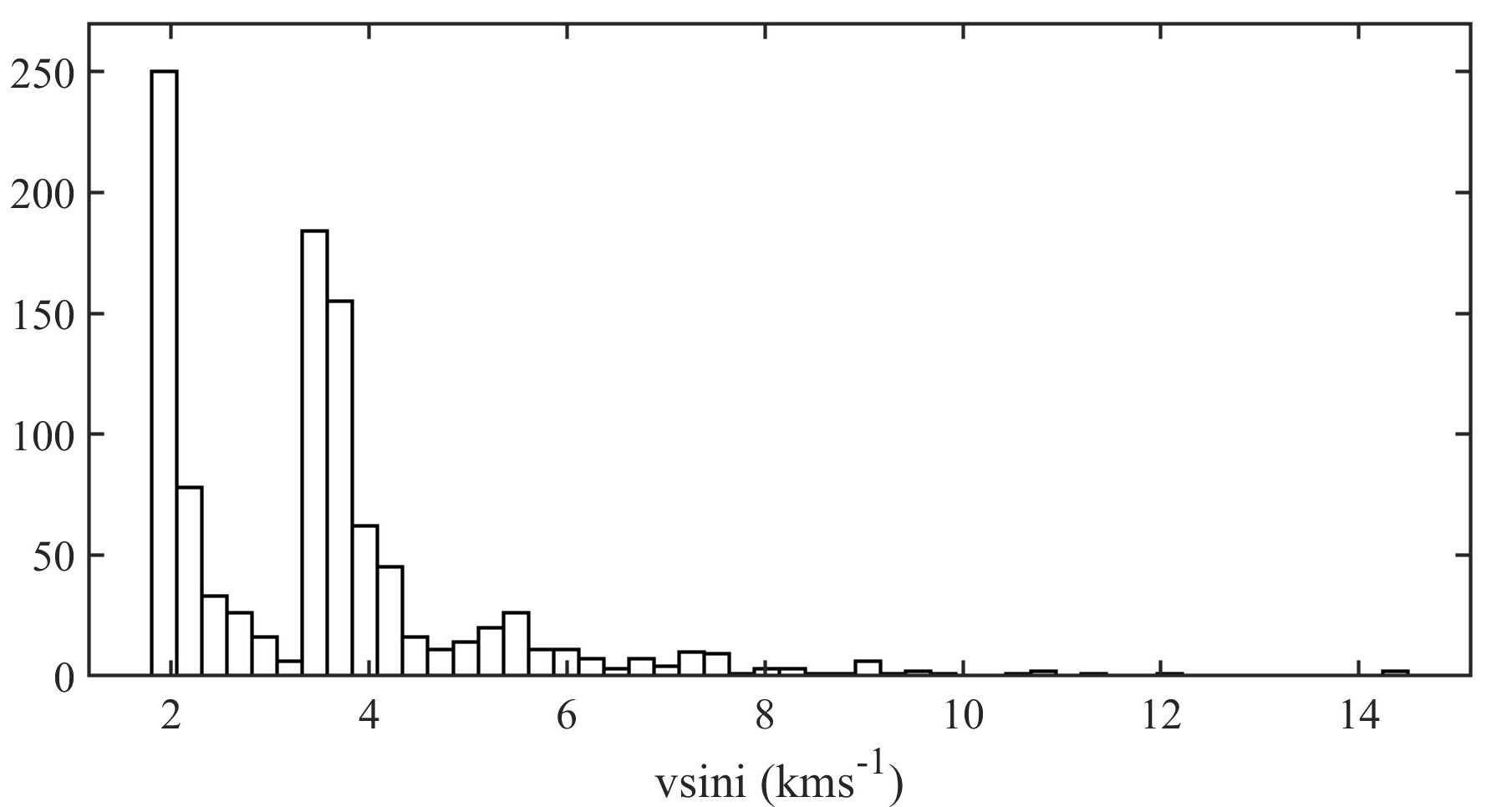}
\end{subfigure}
\begin{subfigure}{.45\textwidth}
\includegraphics[width=\linewidth]{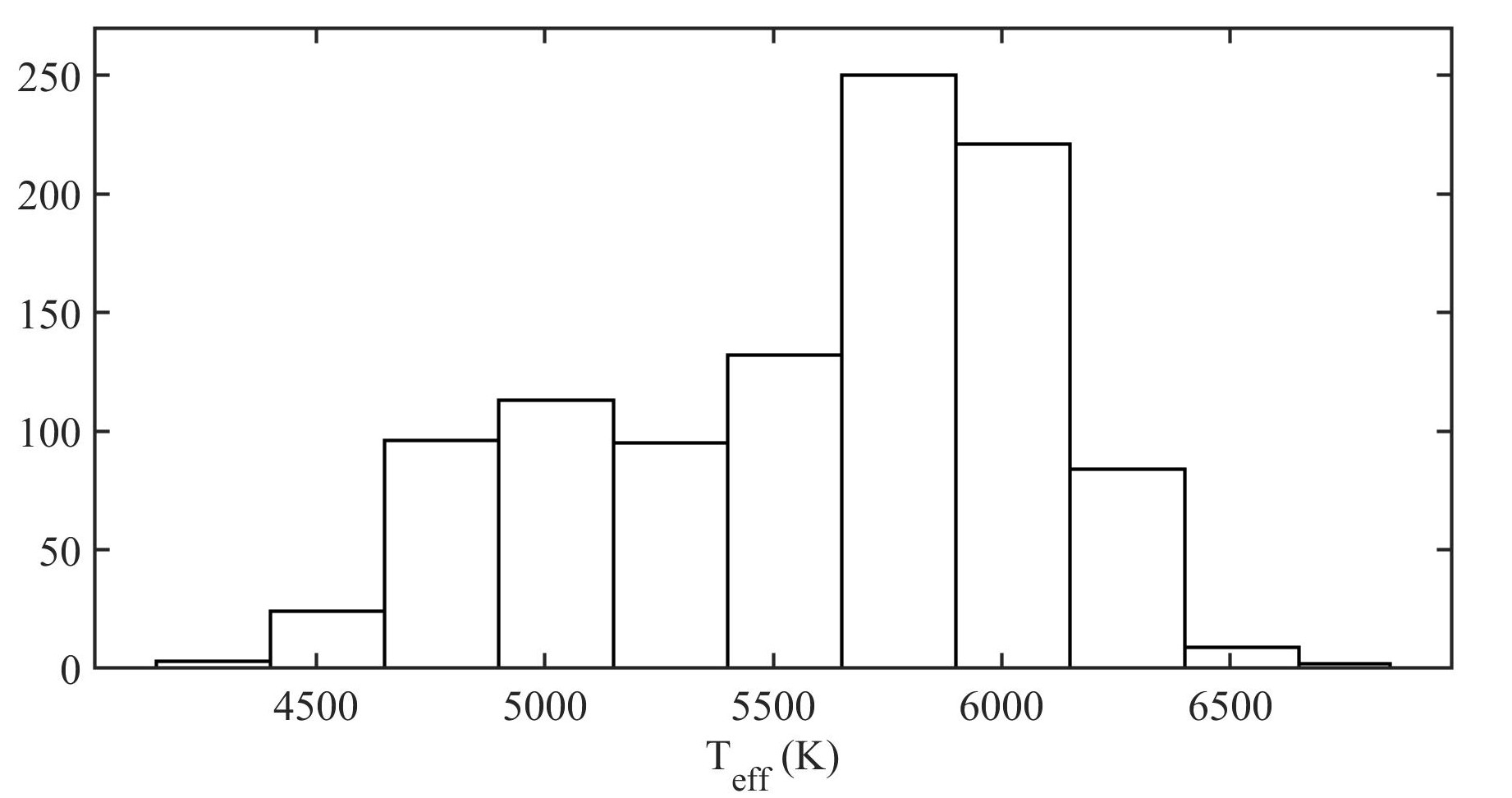}
\end{subfigure}
\caption{Left panel: Distribution of rotation velocities in our sample. Rotation velocities of stars with $vsini < 2\, km\,s^{-1}$ are at the limit of possible measurement. Right panel: Distribution of effective temperatures in our sample.}
\label{fig:teffdist}
\end{center}
\end{figure}

This sample provides an excellent foundation for studying stellar rotation and magnetic activity for several key characteristics. It comprises a well-characterised population of bright, slowly rotating FGK-type stars—ideal targets for the accurate determination of both rotation periods and magnetic cycle periods through long-term monitoring. As shown in the left panel of Fig. \ref{fig:teffdist}, the distribution of projected rotational velocities (vsini) estimated by \cite{boulkaboul2022analysis} has a median value of $3.515\pm 0.051\, kms^{-1}$, confirming that these stars are predominantly slow rotators. The concentration near $vsini \approx 1.5 \, km\,s^{-1}$ corresponds to the minimum detectable value, where rotational broadening becomes comparable to other line–broadening mechanisms such as instrumental profile and macroturbulence, making slower rotations indistinguishable. The broader peak at $3.5 \, km\,s^{-1}$ reflects the typical projected rotation of slow-rotating solar-type main-sequence stars.
Right panel of figure \ref{fig:teffdist} displays the effective temperature distribution derived by \cite{boulkaboul2022analysis}, with a median around $5750\pm 20 K$, consistent with late-type main-sequence stars. These stars are nearby, with a median distance of $45.2 \pm 0.5$ pc \citep{bailer2021estimating}, and benefit from high signal-to-noise photometric and spectroscopic data from facilities such as HARPS and TESS, which enable detailed time-series analyses. Furthermore, many of these stars have been observed over decades in photometric and chromospheric activity surveys, making them particularly well suited for detecting long-term magnetic cycles. The sample is also well-suited for this study as it excludes in principle binary and multiple star systems, thereby reducing potential complications from stellar companions.

Importantly, the sample includes both stars with and without confirmed exoplanet companions. Specifically, 140 stars from a total of 1031 in the sample have reported planet detections based on either the RV or transit method, as listed in the NASA Exoplanet Archive\footnote{\url{https://exoplanetarchive.ipac.caltech.edu/}} and the Exoplanet database\footnote{\url{http://exoplanet.eu/}}. This mix allows us to explore potential correlations between the presence of planets and stellar rotation or magnetic activity properties. Moreover, the homogeneous selection criteria and consistently high-quality data across the catalogue help reduce observational biases and avoid mixing of possibly different dynamo types that often affect broader stellar samples, thereby enabling a more robust comparison between planet-hosting and non-planet-hosting stars.

\section{METHODS}
\label{method}

\subsection{Radial velocity measurement}
The RVs used in this work are those determined by \citep{boulkaboul2022analysis}. They were derived by matching archived HARPS spectra with theoretical templates derived from the MARCS \citep{gustafsson2008grid} and AMBRE \citep{de2012ambre} atmosphere models, which are available in the Pollux database\footnote{\url{http://pollux.oreme.org}}. This matching process minimises the quadratic sum, $\chi^2$, of the differences between the observed and theoretical spectra. The methodology, as detailed in \cite{boulkaboul2022analysis}, can be summarised as follows: 

The synthetic spectrum $s(\lambda_s)$, defined on its rest-frame wavelength grid $\lambda_s$, is first resampled onto the observed wavelength grid following the formalism of \citet{david2014multi}. The template is then convolved with the instrumental line-spread function, assumed to be Gaussian, and with a rotational broadening kernel $u(v\sin i)$ following \citet{gray2021observation}. The resulting synthetic spectrum is Doppler shifted by a trial radial velocity RV, yielding
\begin{equation}
\label{resample}
T(\lambda, \mathrm{RV}, v_{\rm vsini}) = u(v_{\rm vsini}) * G\left[\lambda, \lambda_s\left(1 + \frac{\mathrm{RV}}{c}\right)\right] * s(\lambda_s),
\end{equation}
where $G$ denotes Green's function convolved with a Gaussian kernel. The RV is obtained by minimizing the weighted quadratic difference between the observed spectrum $S_i$ and the synthetic spectrum $T'_i$:
\begin{align}
\chi^2 &= \sum_i w_i \left(S_i - T'_i\right)^2, \
\label{sb1}
T'i &= P_n(\lambda_i), T(\lambda_i, \mathrm{RV}, v{\rm vsini}) + b,
\end{align}
where $w_i$ are the statistical weights, $P_n(\lambda)$ is a polynomial accounting for the continuum, and $b$ represents the background flux.
We iterated over both plane-parallel and spherical MARCS-AMBRE synthetic spectra, varying the initial atmospheric parameters (APs)—effective temperature $T_{\mathrm{eff}}$, surface gravity $\log g$, and metallicity [Fe$\slash$H]—within the intervals [$p - \Delta \mathrm{p}, p + \Delta \mathrm{p}$], where $p$ represents one of the APs and $\Delta \mathrm{p}$ is the respective step size. The starting values of these parameters were obtained from \cite{soubiran2016pastel}. The step sizes are set as $\Delta T_{\mathrm{eff}} = 250$K, $\Delta \log g = 0.5$ dex, and $\Delta$[Fe$\slash$H] = 0.25 dex.

If any of the parameters reached the boundary of their defined interval $p_0 = [p \pm \Delta \mathrm{p}]$, a second iteration was carried out, extending the parameter range to $[p, p + 2 \times \Delta \mathrm{p}]$ or $[p - 2 \times \Delta \mathrm{p}, p]$, as needed. Convergence was achieved when the parameters settled within the centre of the defined range. This iterative process allowed us to measure the star's RVs and select the best-fitting template with the atmospheric parameters and projected rotation velocity (vsini). The detailed parameters for the targets can be found in the CDS\footnote{\url{VizieR DOI : https://doi.org/10.26093/cds/vizier.75171849}} of \cite{boulkaboul2022analysis}. An example for the fitted spectrum of HIP12653 is provided in Figure 2 of \cite{boulkaboul2025magnetic}, which shows a fitted HARPS spectrum with overplotted best-fit synthetic spectrum in the spectral range $4450-4471 $\AA.

Across the entire sample, the mean internal error of the RV measurements is $2.1\,\mathrm{m\,s^{-1}}$. The root mean square (RMS) of the raw RV time series, computed before subtracting any planetary signals, and therefore including contributions from both orbital motion and stellar variability, has a mean value of $21\,\mathrm{m\,s^{-1}}$ for stars with planetary companions (reaching up to $133\,\mathrm{m\,s^{-1}}$), and $6\,\mathrm{m\,s^{-1}}$ for stars without companions (with a maximum of $85\,\mathrm{m\,s^{-1}}$). These relatively high RMS values likely include contributions from undetected RV variability.

\subsection{Magnetic activity indicators}
In stars hosting solar-type dynamos, the magnetic field induces activity in the chromosphere and photosphere, which is related to the convection zone structure, the star’s rotation, and magnetic field regeneration \citep{hall2008stellar}. This activity manifests itself through surface features such as stellar spots and plages, which, due to stellar rotation, cause RV variations (RV jitter). For low-activity stars with slow rotation, RV jitter is typically under $5 \,\rm{m \ s^{-1}}$, but can reach $1000 \,\rm{m \ s^{-1}}$ in highly active stars \citep{weise2010search}.  

\subsubsection{Line profile indicator}
RV variations caused by stellar atmosphere changes are reflected in spectral line profile asymmetries, which can be measured using the line bisector technique. This technique helps distinguish RV jitter from signals due to companion stars, as bisectors oscillate around a mean value without changing shape or orientation when caused by a companion \citep{queloz2001no, motalebi2015harps}. Among the magnetic activity indicators that quantify variations in the shape of the spectral line bisector, we use the Bisector Inverse Slope (BIS), as defined by \citet{queloz2001no} and \citet{figueira2013line}. The BIS is calculated as the difference in velocity between the bisector at the top (the average of the midpoints between 60\% and 90\% of the cross-correlation function (CCF)) and the bottom (the average of the midpoints between 10\% and 40\% of the CCF). The BIS values are extracted directly from the instrument pipeline.

In stellar spectra, two different activity regimes can produce opposite signs of the BIS–RV correlation depending on the underlying physical mechanism and timescale. In the case of RV variation caused by photospheric active regions (spots in rotation), the top of the CCF bisector is much less affected than the lower part, which oscillates around an average value. In this rotational modulation regime, a localised dark spot sweeping across the visible hemisphere suppresses flux asymmetrically, distorting the line profile in a manner that is geometrically out of phase with the associated RV shift, leading to an anti-correlation between BIS and RV at the rotation period (see e.g. \citep{queloz2001no, boisse2011disentangling, da2012long}). In the magnetic cycle regime, the secular growth and decay of the total active region filling factor modulates the disc-integrated convective blueshift suppression and mean line asymmetry in the same direction, producing a coherent positive correlation between BIS and RV on cycle timescales \citep{queloz2001no, hatzes2010investigation, lovis2011harps}.

\subsubsection{Chromospheric activity indicator}
\label{subsect:chromAct}
Spectral lines sensitive to changes in the chromosphere, such as the \ion{Ca}{ii} and \ion{Mg}{ii} lines, are commonly used to quantify chromospheric activity. Since the \ion{Mg}{ii} lines fall outside the wavelength range of the HARPS spectrograph, we only use the \ion{Ca}{ii} lines.

To measure the strength of the magnetic field in the \ion{Ca}{ii} H \& K emission lines, we estimate the S-index using the method described by \citet{lovis2011harps}. The index is computed as the ratio between the flux in the line cores and that in two nearby continuum passbands. Specifically, we sum the flux within two triangular bandpasses of width 1.09 \AA\ centered on the K ($3933.66$ \AA) and H ($3968.47$ \AA) lines, and divide by the sum of the flux in two reference windows of 20 \AA\ width centered on the continuum passbands V ($3900.0$ \AA) and R ($4000.0$ \AA), where the wavelengths are in air. These passbands are defined consistently with those used by the Mount Wilson HKP-2 spectrometer \citep{duncan1991ii}. The S-index is given by:

\begin{equation}
\mbox{S-index} = \alpha \frac{H + K}{R + V}
\end{equation}
where the calibration constant $\alpha$ is fixed at 2.4, as specified by \citet{duncan1991ii}. The resulting HARPS S-index values are then calibrated onto the Mount Wilson scale using $S_{\rm MW} = 1.111 \times S_{\rm HARPS} + 0.0153$ following \cite{lovis2011harps}. We use HARPS 1D spectra, which are blaze-corrected by the pipeline but neither flux-calibrated nor continuum-normalized. Since the S-index is defined as a ratio of fluxes measured within a very narrow wavelength range, absolute flux calibration is not required, and the index remains consistent across the dataset.

The HARPS spectra wavelength scale is transformed to the rest frame of the star using the measured RV before computing the S-index.

A correlation between RV and the S-index over a magnetic cycle is explained by the local inhibition of convection in magnetically active regions, where a stronger magnetic field reduces convective motions. This reduces the blue shift in the active regions, causing the spectral lines to shift towards the red when the magnetic field is strong. 

If the period of variation in the magnetic activity indices matches that of the RV variation or one of its harmonics, it suggests that the RV changes are likely due to inhomogeneities in the stellar atmosphere. However, if the RV period does not align with the magnetic activity indicators period, it does not rule out the possibility that the variations stem from a magnetic origin \citep{santos2010stellar}.

\subsection{Period search}
The period search in the RV and activity indicators datasets is based on the periodogram method of \citet*{heck1985period} (HMM), as revisited in \cite{zechmeister2009generalised} using the Generalized Least Squares (GLS) approach. To assess the significance of detected periods, we construct an empirical cumulative probability distribution function (ECDF) from the highest peaks of $10^6$ realisations of white noise periodograms with the same time sampling. In this framework, our adopted $95 \%$ probability threshold corresponds to a false alarm probability (FAP) of 0.05. This relatively generous threshold is chosen to allow the identification of potential periodicities even in stars with limited numbers of spectra, where applying stricter detection thresholds (FAP $ < 0.01$) might lead to the rejection of real but low-significance signals. The uncertainties associated with the fitted orbital parameters, including the period, are derived from the variance–covariance matrix, following the procedure described in \citet[Section 15.4.1]{press1986numerical}. 
The variance of the measured frequency is inversely proportional to both the time baseline and the number of measurements; consequently, the uncertainty on the derived period scales proportionally with the period itself, hence longer time series and larger datasets lead to smaller uncertainties. In contrast, most rotation periods reported in literature are inferred from empirical activity–rotation relations \citep{mamajek2008improved}, while long-term cycle periods are estimated using bootstrap GLS analyses based on the $68\%$ confidence interval of the distribution of peak periods \citep{lovis2011harps}. This procedure probes the stability of the detected period against short-term variability (e.g. active-region evolution and differential rotation) rather than the formal fitting precision. As a result, the quoted uncertainties in those studies are expected to be significantly larger than our estimated errors. We note that the period uncertainties reported here are computed from the rescaled RV errors.
Following \citep{halbwachs2023gaia}, we apply Wilson–Hilferty transformation and use $F_2$ goodness-of-fit statistic, $F_2 = \sqrt{\frac{9\rm{DoF}}{2}}\left[\left(\frac{\chi^2}{\rm{DoF}}\right)^{1/3} + \frac{2}{9\rm{DoF}}-1\right]$, where $\chi^2$ is the sum of the squared normalized residuals and DoF is the number of degrees of freedom. For an adequate model with correctly estimated uncertainties, $F_2$ is expected to follow a normal distribution $\mathcal{N}(0,1)$. We therefore rescale the RV error bars by a factor $c = \sqrt{\frac{\chi^2}{\rm{DoF}\left(1 - \frac{2}{9\rm{DoF}}\right)^3}}$, so that the resulting $F_2$ is equal to zero.

For the RV dataset, the dominant periodicities are identified from the GLS periodogram as the frequencies corresponding to the highest peak in each iteration, provided that their power exceeds the adopted threshold. These periodicities may correspond to planetary signals, stellar rotation, or magnetic activity cycles. The primary signal's best-fitting Keplerian solution is obtained using the algorithm of \citet{zechmeister2009generalised}. The derived model is then subtracted from the original RV time series to obtain residuals, to which we apply the GLS again to search for additional periodicities. Since signals from derived periods are not orthogonal, a global refinement is needed; after each newly detected period, a global multi-period Keplerian fit is performed, simultaneously refining all orbital parameters identified up to that point. The process is repeated iteratively—detecting a new signal, updating the global fit, and recomputing the residuals—until no significant peaks remain above the adopted significance threshold.

A similar iterative procedure is applied to the S-index and BIS datasets to identify both rotation and magnetic cycle periods. In each case, the dominant signal is subtracted before recomputing the periodogram on the residuals. As in \cite{lovis2011harps}, a two-year threshold is used to distinguish between the short-term rotation period and the long-term activity cycle period.

While fitting activity indicators with Keplerian model may introduce spurious signals, it remains more appropriate than enforcing a strictly circular model when describing stellar activity cycles. This is because magnetic cycles are not perfectly periodic or sinusoidal \citep{olah2009multiple}. For example, the solar cycle rises faster than it falls, leading to a cycle shape that deviates from a sine function. To model the temporal evolution of the S-index, $S(t)$, it is therefore essential to allow for such asymmetry. We do so by adopting a functional form inspired by the Keplerian radial-velocity equation, $V = V_0 + K[e cos\omega + cos(\omega + \nu)]$. We apply this model to fit the S-index time series, $S(t) = S_0 + A[e \cos\omega + \cos(\omega + \nu)]$, adjusting the orbital parameters to describe features of the magnetic activity cycle rather than orbital motion. In this model, the eccentricity, $e$, quantifies the degree of asymmetry in the modulation, representing how the rise and fall of magnetic activity deviate from a purely sinusoidal variation. Specifically, it indicates whether one phase (either rising or decaying) is more prolonged than the other, as detailed in \cite{boulkaboul2025magnetic}. While this parameterization allows for asymmetric cycle shapes, it assumes that each detected periodic component is locally coherent over the time span in which it is fitted, with a well-defined period, amplitude, and asymmetry. Multiple magnetic cycles are handled by iteratively fitting significant periodicities and modeling each with its own component, allowing different cycles to have different asymmetries. For stars with complex or multi-periodic activity (e.g., LQ Hya), this approach represents the observed variability as a superposition of several asymmetric cycles. However, if a single physical cycle evolves continuously in time, its period, amplitude, or rise–decay asymmetry drift significantly from one epoch to another, then a single component should be interpreted as a local or effective description of that cycle over the observing window.

A period search is also performed on the photometric data to confirm potential new rotation period estimates derived from the S-index and/or BIS datasets. The associated period uncertainties are computed using the flux errors provided by the TESS reduction pipeline.

\section{Results}
\label{results}
\begin{figure}
\begin{subfigure}{.5\textwidth}
\includegraphics[width=\linewidth]{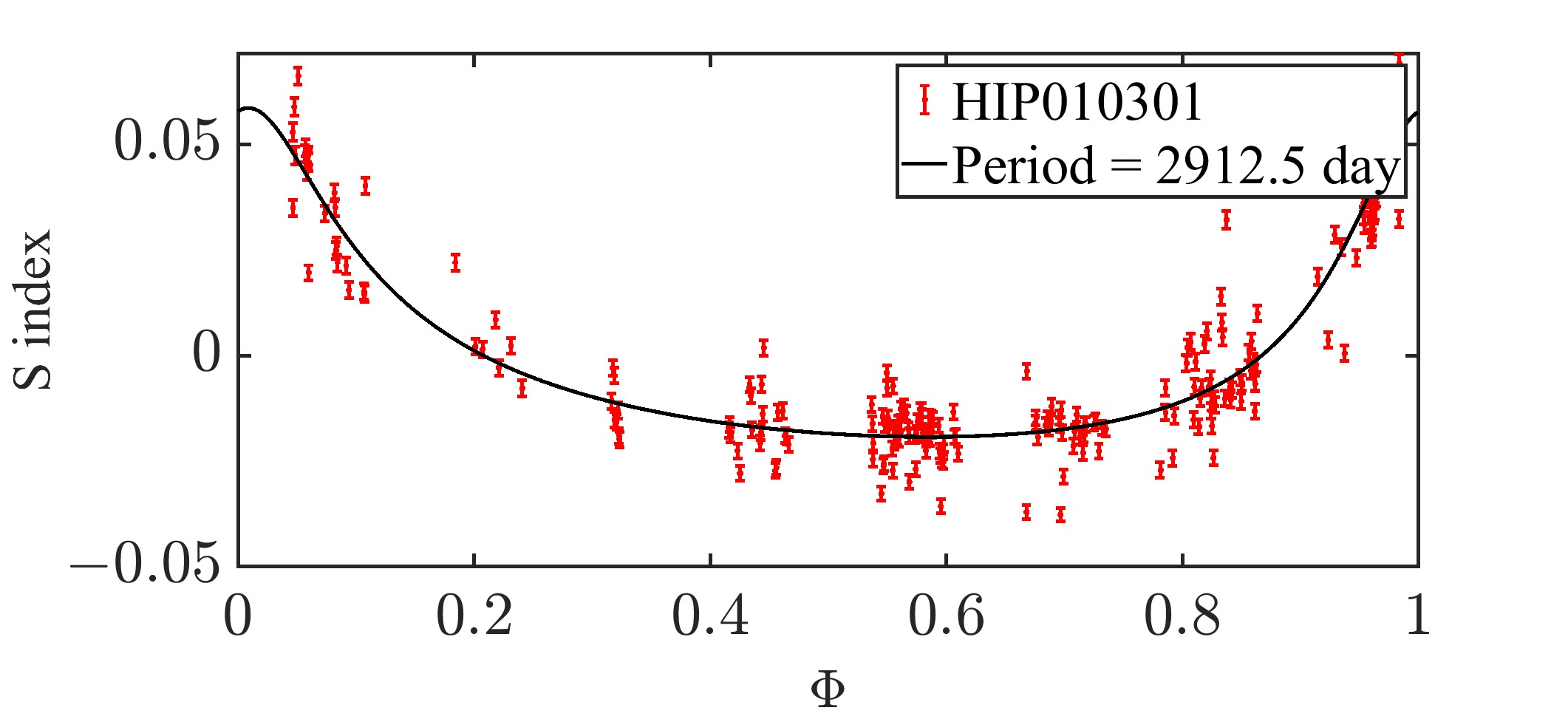}
\end{subfigure}
\begin{subfigure}{.5\textwidth}
\includegraphics[width=\linewidth]{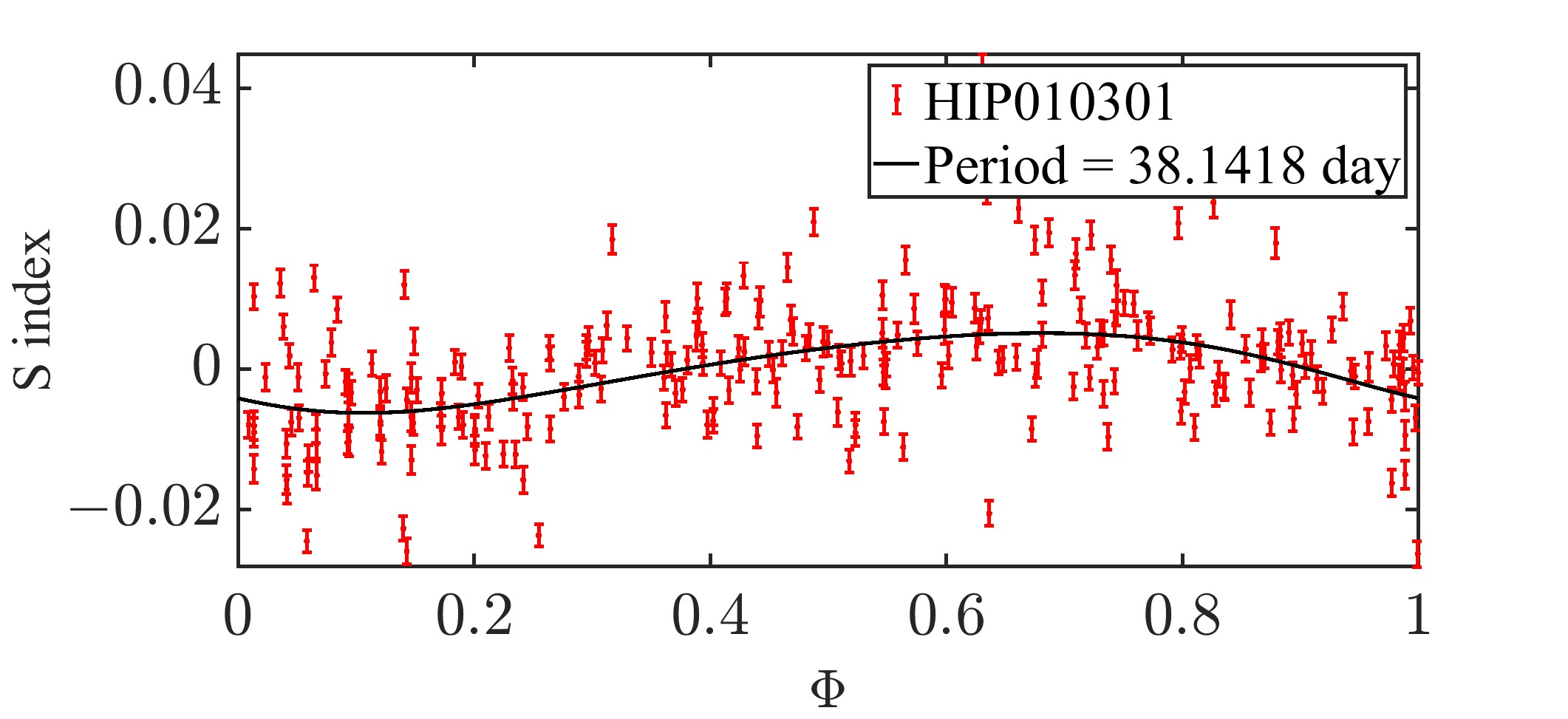}
\end{subfigure}
\caption{Phase-folded S-index data fitted with the long period corresponding to the magnetic cycle (left panel) and short period corresponding to the rotation period (right panel).}
\label{fig:magProtHIP10301}
\end{figure}

Out of the 1031 stars in the HARPS catalogue, 767 have a sufficient number of measurements (more than ten observations) to allow for a reliable period search. Figure \ref{fig:magProtHIP10301} shows an example of the S-index time series for HIP10301, phase-folded over the magnetic cycle (left), where the asymmetry in the cycle is clear ($e = 0.63$), and rotation periods (right), using the best-fit values.

\subsection{Activity periods in RV dataset}
\subsubsection{Stars with planetary companions}
Among the 134 stars with more than ten spectra, 127 have a significant period detection in the RV dataset. We list in Table \ref{tab:RV_comp} the stars with confirmed planetary companions for which rotation and magnetic cycle periods extracted from literature are close to periods detected in the RV dataset within three times the uncertainties, along with those detected in S-index and BIS. We include those close to half of the rotation period, such as HIP70695, HIP93540, and HIP108375. Below, we discuss the periods assigned to planetary companions:
\begin{itemize}
    \item \textbf{HIP28460}. A G2V-type \citep{perryman1997hipparcos}, $2^{+3}_{-2}$ Gyr old star \citet{jenkins2013two}, with two planetary companions with periods of $18.357^{+0.066}_{-0.044}$ and $25.648^{+0.12}_{-0.25}$ d \citep{jenkins2013two}.
    The $25.648^{+0.12}_{-0.25}$-day period closely matches the $26.4\pm 1.1$-day rotation period identified by \citet{suarez2015rotation}. In addition to the period of $34.954 \pm 0.022$ d, we also detect another period of $69.582 \pm 0.053$ d in the S-index periodogram. This is consistent with the peak of 69 days detected in the S-index periodogram of \citet{jenkins2013two}, which they suggested may represent the star's true rotation period. Additionally, we observe a statistically significant positive correlation between the RV and BIS, with a p-value of $0.0088$, shown in the middle-left panel of Fig. \ref{fig:correlation}. These findings favor a rotational rather than a planetary origin for the signal.

    \item \textbf{HIP40693}. A K0V-type, approximately 4.47 Gyr old star \citep{turnbull2015exocat}, hosting three planets with periods of $8.667\pm0.003 $ d, $31.56\pm0.04$ d, and $197\pm3 $ d \citep{lovis2006extrasolar}. Our estimated planetary period of $31.6132 \pm 0.0091$ d is close to both our S-index period ($33.8442 \pm 0.0022$ d) and the stellar rotation period ($35.8 \pm 3.9$ d) derived by \citet{lovis2006extrasolar} from activity indicators. Although they noted that the BIS does not show a signal of comparable amplitude to the RV variation, this does not exclude a stellar origin. In our analysis, we find a significant positive correlation (p-value $\leq 5\%$) between the RV and both the BIS and S-index, suggesting a possible connection to stellar rotation. The RV variation versus S-index is shown in the middle-right panel of Fig. \ref{fig:correlation}.

    \item \textbf{HIP48331}. A K5V-type, $5.61\pm0.61$ Gyr old star, hosting a planet with a mass of $0.011\pm0.002\, M_J$ and an orbital period of $58.43 \pm 0.13  $ d \citep{pepe2011harps}. Our RV period search reveals a signal at $51.41 \pm 0.24$ d, which is close to both the estimated stellar rotation period of $47.1 \pm 7.0 $ d \citep{pepe2011harps} and the S-index periodicity of $ 47.3847 \pm 0.0013$ d, suggesting a stellar rather than planetary origin. This interpretation is consistent with \citet{laliotis2023doppler}, who attributed the RV variability to stellar rotation instead of a planetary companion. This is further supported by the significant negative correlation between RV and the S-index, shown in the bottom-right panel of Fig. \ref{fig:correlation}. The lack of correlation with the BIS is consistent with the absence of a corresponding periodicity in the BIS data.

    \item \textbf{HIP54400}. A G0V-type, $8.6\pm0.7$ Gyr old star, hosting three planets orbiting with periods of $8.1245\pm0.0006$ d, $19.88\pm0.01$ d, and $103.5\pm0.1$ d \citep{unger2021harps}.
    Our RV period of $19.888 \pm 0.018$-day is close to the stellar rotation period of $20.6\pm 2.9$ days \citep{lovis2011harps}, but we do not detect this period in the activity indicators. \citet{unger2021harps} suggested that the $19.88 \pm 0.01$-day signal is a harmonic of the rotation period, resulting from active regions on opposite hemispheres of the star. They proposed that the 41-day period observed in $\log R'_{\rm{HK}}$ corresponds to the star's rotation period. This interpretation is plausible, as our BIS periodogram shows a signal at \(46.45 \pm 0.15\) d, and the S-index periodogram yields a consistent period of \(46.617 \pm 0.049\) d. Furthermore, we find a significant correlation between the BIS and RV, shown in the bottom-left panel of Fig. \ref{fig:correlation}, supporting a stellar origin for the observed variability.

     \item \textbf{HIP57931}. A K1V(P)-type\footnote{\url{https://simbad.cds.unistra.fr/}}, 9.1 Gyr old star, with an intermediate-mass planet orbiting with a period of $47.84 \pm 0.03$ d \citep{mordasini2011harps}.
     \citet{mordasini2011harps} noted the similarity between this RV signal and an estimate of the stellar rotation period ($51 \pm 5$ days) derived from an empirical activity-rotation relation. However, they did not attribute the RV signal to stellar rotation since this period was not observed in their activity diagnostics. Similarly, this period was not detected in our activity indicators. Moreover, we find no significant correlation between the RV and either the S-index or BIS, suggesting that the signal is unlikely to be related to stellar activity.

     \item \textbf{HIP93373}. A G6V-type, $10.32\pm1.58$ Gyr old star, hosting a planet with an orbital period of $29.01 \pm 0.02$ d \citet{mortier2016harps}. \citet{mortier2016harps} argued that this RV signal, although close to the stellar rotation period of $28.95 \pm 0.33$ d, is consistent with the presence of a sub-Neptune companion. Our results support a planetary interpretation, as we do not detect any corresponding signal in the activity indicators, and no significant correlation is found between the RV and either the S-index or BIS.

     \item \textbf{HIP106006}. A G5V-type, $3.38\pm2.58$ Gyr old star \citep{segransan2011harps}, with three planets confirmed to orbit it with periods of $34.905\pm0.012$ d, $2024.1\pm3.1$ d \citet{diaz2016harps}, and $7326^{+400}_{-369}$ d \citep{feng20223d}. Similarly to \citet{diaz2016harps}, the rotation period of $34.1 \pm 3.7$ days, which they estimated from the $\log(R'_{\mathrm{HK}})$--rotation relation of \citet{mamajek2008improved}, is not present in our BIS periodogram as well, and there is no significant correlation between the RV and either activity indicators. We note that \citet{diaz2016harps} also reported a period of $\sim$29.5 days in the FWHM GLS periodogram, which is closer to our S-index period of $26.8804 \pm 0.0013$ d than to our RV signal of $34.955 \pm 0.025$ d. Taken together, these results support a planetary origin for the $\sim$35-day RV signal.

     \item \textbf{HIP118319}. A G2V-type, $4.12\pm0.67$ Gyr old star, with a planetary companion orbiting with a period of $26.6904\pm0.0019$ d \citet{ment2018radial}. Although the orbital and rotation periods are similar within uncertainties, \citet{johnson2006n2k} reported no photometric variability at the RV period of $26.73 \pm0.02$ d. In our analysis, the period detected in the S-index dataset, $ 22.95 \pm 0.35$ d, is closer to the additional RV period of $24.689 \pm 0.010$ days ($K = 9.6 \pm 2.4 \, m s^{-1}$), and there is no significant correlation between the RV and either the S-index or BIS. These results support a planetary origin for the RV signal of $26.7$ d.

 \end{itemize}
\begin{table}
\begin{center}
\tiny
    \caption{\small The rotation periods ($P_{\rm{rot}}$) and the magnetic cycle periods ($P_{\rm{mag}}$) are presented in the same column but are separated by a horizontal line. The table includes literature values for both the rotation and magnetic cycle periods (Lit $P_{\rm{rot}}$/ Lit $P_{\rm{mag}}$), along with periods derived from S-index ($P_{\rm{S}}$) and BIS ($P_{\rm{BIS}}$) periodograms close to RV periods ($P_{\rm{RV}}$) with RV semi-amplitude ($K_{\mathrm{RV}}$) for stars with confirmed planetary companion. The number of spectra, $N$, is listed in the last column.}
    \vspace*{5mm}
    \label{tab:RV_comp}
    \begin{tabular}{rrrrrrrr}
    \hline
HIP & Lit $P_{\rm{rot}}/P_{\rm{mag}}$ [d] & Ref. & $P_{\rm{S}}$ [d] & $P_{\rm{BIS}}$ [d] & $P_{\rm{RV}}$ [d] &$K_{\mathrm{RV}} [m s^{-1}]$ & N\\
\hline
1499 & $30.2 \pm 3.5$  & \protect\cite{diaz2016harps}& $13.5506 \pm 0.0020$ & -- &$13.5065 \pm 0.0038$ & $1.80 \pm 0.26$ &467 \\
7978 & $10 \pm 3$ &\protect\cite{marmier2013coralie} & $11.05728 \pm 0.00024$& $11.040279 \pm 0.000064$& $11.5830 \pm 0.0018 $ &$3.20 \pm 0.50$ & 195\\ 
11433 & $36.528 \pm 0.022$ &\protect\cite{gandolfi2019transiting}  &  $37.684 \pm 0.011$ &  --&  $37.222 \pm 0.016$ & $3.01 \pm 0.89$& 116\\ 
12186 & $27.8 \pm 3.2$ &\protect\cite{lovis2011harps} &  $23.9544 \pm 0.0096$ & -- &  $32.697 \pm 0.027$&$1.78 \pm 0.32$ & 296\\
12653 & $7.92 \pm 1.63$ &\protect\cite{watson2010estimating} &  $4.76952 \pm 0.00023$& $5.90363 \pm 0.00010$ &  $5.715924 \pm 0.000060$ & $10.71 \pm 0.25$& 2118\\
15510 &$33.19 \pm 3.61$ &\protect\cite{pepe2011harps}& $88.6208 \pm 0.0034$ &  --&$89.402 \pm 0.067$&$0.706 \pm 0.087$ &6782\\
26013 & $30.5 \pm 0.7$ &\protect\cite{osborn2021toi}  &    $29.10497 \pm 0.00087$  &  $27.4307 \pm 0.0047$&  $28.5817 \pm 0.0067$ &$5.31 \pm 0.53$ & 172\\
28460 & $26.4 \pm 1.1$ &\protect\cite{suarez2015rotation} &$34.954 \pm 0.022$ & -- & $25.6401 \pm 0.0097$ &$3.28 \pm 0.39$ & 226\\
30503 & $21.4 \pm 3.0 $ &\protect\cite{boro2018chromospheric}&  $19.9216 \pm 0.0027$ &  -- & $20.2066 \pm 0.0028$ & $2.70 \pm 0.44$& 324\\
36795 & $7.0 \pm 1.0$ &\protect\cite{desort2008extrasolar} &  -- &  -- &  $6.3014 \pm 0.0011$ & $4.2 \pm 1.9$& 195\\
37284 & $20.3 \pm 5.3$ &\protect\cite{watson2010estimating} &  -- &  -- &  $18.2702 \pm 0.0056 $ & $10.0 \pm 2.3$& 41\\
40693 & $35.8 \pm 3.9$ &\protect\cite{lovis2011harps} & $33.8442 \pm 0.0022$& $40.5465 \pm 0.0045$& $31.6132 \pm 0.0091$ & $2.62 \pm 0.17$& 741\\
44291 & $33  \pm 10$ &\protect\cite{suarez2015rotation} &  -- &  -- &  $28.2024 \pm 0.0045$ & $3.93 \pm 0.97$& 160\\
47007 & $21.9 \pm 1.9 $ &\protect\cite{watson2010estimating} &  -- &  $27.902 \pm 0.005$ &  $17.7187 \pm 0.0038$&$3.23 \pm 0.47$ & 255\\
48331 & $47.1 \pm 7.0 $ &\protect\cite{pepe2011harps}& $ 47.3847 \pm 0.0013$& $34.28 \pm 0.15$&  $51.41 \pm 0.24$ &$0.42 \pm 0.22$& 1191\\
54400 & $20.6 \pm 2.9 $ &\protect\cite{lovis2011harps}&  $46.617 \pm 0.049$ &  $46.45 \pm 0.15$ &  $19.888 \pm 0.018$&$1.00 \pm 0.21$ & 362\\
57370 & $12.3 \pm 0.20$ &\protect\cite{ge2006first} &  $6.679 \pm 0.012$ & --& $7.9337 \pm 0.0041$ & $14.8 \pm 1.8$& 24\\
57931 & $51.0  \pm  5.0 $ &\protect\cite{zoghbi2011quantization}&  -- &  -- &  $47.723 \pm 0.049$ & $5.52 \pm 0.86$& 64\\
70695 & $30.3  \pm 3.4$ &\protect\cite{lovis2011harps} &  -- &  $16.0222 \pm 0.0012$ &  $16.927 \pm 0.014$ & $1.15 \pm 0.38$& 146\\
93373 & $28.95 \pm 0.33$ &\protect\cite{mortier2016harps} &  -- &  -- &  $29.040 \pm 0.030$&$2.24 \pm 0.42$&127 \\
93540 & $35.90 \pm 0.20$ &\protect\cite{mascareno2018ropes}   & $33.3594 \pm 0.0039$&--&  $16.7997 \pm 0.0061$ &$2.48 \pm 0.46$& 169\\
99825 & $47.7 \pm 4.9$ &\protect\cite{pepe2011harps}  &  $48.935 \pm 0.037$& $45.870 \pm  0.019$ &   $ 47.580 \pm 0.022$ &$1.00 \pm 0.30$& 1796\\ 
106006 & $34.1 \pm 3.7$ &\protect\cite{diaz2016harps} &  $26.8804 \pm 0.0013$ &  -- &  $34.955 \pm 0.025$&$2.94 \pm 0.48$ & 112\\
107985 & $17.8  \pm 0.5$ &\protect\cite{haghighipour2012lick} &  -- &  $20.334 \pm  0.027 $&  $19.9693 \pm 0.0087$ &$6.56 \pm 0.92$& 73\\
108375 & $12.4 \pm 1.1$ &\protect\cite{watson2010estimating} &  -- &  -- &  $6.73504 \pm 0.00086$ &$2.14 \pm 0.62$& 126\\
113044 & $21.3 \pm 1.6$ &\protect\cite{watson2010estimating} &  -- &  $15.611 \pm 0.005$ &  $21.1797 \pm 0.0031$ &$7.2 \pm 1.4$& 167\\
113238 & $35.6  \pm 1.0$ &\protect\cite{mayor2004coralie} &  -- &  $31.6192 \pm 0.0028$ &  $38.558 \pm 0.026 $& $18 \pm 19$&128\\
116084 & $5.0 \pm 2.0$ &\protect\cite{naef2007harps}  & $5.015 \pm 0.014$ & --& $5.25286 \pm 0.00091$&$11.0 \pm 1.6$ & 31\\
118319 & $29.7 \pm 1.5$ &\protect\cite{watson2010estimating}  &  $ 22.95 \pm 0.35$ &  -- &  $26.6963 \pm0.0063$ &$36.0 \pm 1.2$& 83\\
118319 & -- &--  & -- &  -- &  $24.689 \pm 0.010$ & $9.6 \pm 2.4$& 83\\
\hline 
1499 & $3521 \pm 77$ &\protect\cite{diaz2016harps}& $3616 \pm 45$& $4308.6 \pm 8.4$ &$3821 \pm 116$ & $4.12 \pm 0.80$& 467\\
5529 & $1461 \pm 110$ &\protect\cite{boro2018chromospheric}& $2697 \pm 16$& $2815 \pm 16$&$2630 \pm 42$&$11.78 \pm 0.55$ &123 \\
7240 & -- & --& $506.2 \pm 3.4$  & $458.06 \pm 0.98$ & $497.6 \pm 3.2$& $15.95 \pm 0.92$ & 77\\
10301 & $3715 \pm 807$ &\protect\cite{lovis2011harps}& $2912.5 \pm 4.5$& --&$2935 \pm 63$&$2.67 \pm 0.68$ & 245\\
12186 & -- &--& $4671 \pm 431$&--&$1871 \pm 150$ &$ 1.39 \pm 0.31$& 296\\
15510 & $751  \pm 290$ &\protect\cite{lovis2011harps}& $1315 \pm 29$& $2031.38 \pm 0.22$&$1692 \pm 64$&$0.416 \pm 0.054$ &6782\\
16085 & $2703 \pm 110$ &\protect\cite{boro2018chromospheric}& $3749 \pm 41$& $4415 \pm 18$&$3647 \pm 255$&$5.12 \pm 0.42$ & 218\\
27435 & $3406 \pm 570$ &\protect\cite{lovis2011harps}& $2994.44 \pm 0.11$& $4287 \pm 272$ & $2819 \pm 84$ &$3.00 \pm 0.33$ & 252\\
30503 & $1790 \pm 110$ &\protect\cite{boro2018chromospheric}& $1950 \pm 65$& $2079.8 \pm 4.0$ &$2008 \pm 39 $&$3.69 \pm 0.29 $ & 324\\
58263 & -- &--& $2613 \pm 22$& -- & $2415.3 \pm 4.6$ &$74.3 \pm 8.7$& 118\\
83541 & -- &--& $1390.8 \pm 5.7$ & -- &$1518 \pm 69$ & $1.2 \pm 1.2$& 269 \\
85017 & -- &--& $3324 \pm 161$ & $3602 \pm 912$ & $3334 \pm 1914$ & $2.89 \pm 0.77$ & 127\\
99825 & $3792 \pm 806$ &\protect\cite{lovis2011harps}& $3604.1 \pm 1.2$& $3611.87 \pm 0.20$ &$3735 \pm 65$ & $3.54 \pm 0.38$& 1796\\
112190 & $2709 \pm 63$ &\protect\cite{delisle2018harps}& $2709.2 \pm 3.2$& $2181.8 \pm 9.2$ & $2682 \pm 38$ & $3.31 \pm 0.51$& 369\\
\hline
\end{tabular}
\end{center}
\end{table}
We also found that some confirmed planet periods are close to the periods present in the activity indicators (S-index/ BIS):

\begin{itemize}
    \item {\bf HIP1499}. A 7.12 Gyr old G5V-type star \citep{turnbull2015exocat} with two confirmed planets, orbiting it with periods of $5.77152\pm0.00045$ d and $13.5052\pm0.0029 $ d \cite{diaz2016harps}. In addition to the period of $23.6035 \pm 0.0076$ d, a signal at $13.5506 \pm 0.0020$ d is detected in our S-index periodogram, close to our RV signal of $13.5065 \pm 0.0038$ d. In addition, \citet{diaz2016harps} noted that their RV period of $13.5052 \pm 0.0029$ d is close to the first harmonic of the stellar rotation period ($30.2 \pm 3.5$ d), suggesting a likely origin in stellar activity. This interpretation is further supported by the positive correlations observed between the RV and both activity indicators. The RV variation versus BIS is shown in the top left panel of Fig. \ref{fig:correlation}.

    \item \textbf{HIP7240}. A $4.3^{+3.0}_{-2.6}$ Gyr old G1V-type star \citep{holmberg2009geneva}. Our RV period of $497.6 \pm 3.2$-day is close to the planetary period of 494 d listed in the Extrasolar Planets Encyclopaedia \citep{schneider2011defining} \footnote{\url{http://exoplanet.eu}} which appears to be unpublished. This period is close to those detected in both our S-index ($506.2 \pm 3.4$ d) and BIS ($458.06 \pm 0.98$ d), and we observe a positive correlation between both activity indicators and the RV. The RV variation versus S-index is shown in the top-right panel of Fig. \ref{fig:correlation}. This indicates that the observed long-term modulation is linked to the star’s magnetic activity cycle. 

    \item \textbf{HIP15510}. A G8.0V-type, 13.5 Gyr old star \citep{turnbull2015exocat}, which hosts four planetary companions: $18.315\pm0.008$ d, $90.309\pm0.184$ d \cite{pepe2011harps}, $147.02^{+1.43}_{-0.91}$ d \citep{feng2017evidence}, and $647.6^{+2.5}_{-2.7}$ d \citep{nari2025revisiting}. Our S-index reveals a period of $28.79 \pm 0.052$ d close to the rotation period of $33.19 \pm 3.61$ d \citep{pepe2011harps} and another period of $88.6208 \pm 0.0034$ d, which, although not overlapping within uncertainties, lies close to this work's RV period of $89.402 \pm 0.067$ d. Although \citet{pepe2011harps} did not detect a corresponding signal in their activity indicators, our analysis shows significant negative correlations between the RV and both the S-index and BIS. This suggests that the signal is more likely related to stellar activity than to a planetary companion.
\end{itemize}

Overall, we confirm that the signals at $25.6401 \pm 0.0097 $d (HIP28460), $51.41 \pm 0.24$ d (HIP48331), and $497.6 \pm 3.2$ d (HIP7240) are attributable to stellar activity. In contrast, the origins of the signals detected at $13.5065 \pm 0.0038$ d (HIP1499), $89.402 \pm 0.067 $d (HIP15510), $31.6132 \pm 0.0091$ d (HIP40693), and $19.888 \pm 0.018$ d (HIP54400) remain ambiguous and warrant further investigation for confirmation. Apart from the three stars to be removed from this subsample, only eight stars have a close planetary period to either rotation or S-index periods. For the remaining stars, those showing no significant correlation between RV and either activity indicator tend to have RV periods that are close only to rotation periods reported in the literature, suggesting that their signals are unlikely to be activity-related. In contrast, all stars with RV periods close to magnetic cycle timescales show significant correlations between RV and at least one activity indicator (S-index or BIS). After reassigning the three excluded targets to the sample of stars without confirmed planetary companions, 35 out of 124 stars show an RV period that is close to at least one activity-related period, namely a period derived from the BIS, the S-index, or a literature value of the stellar rotation or magnetic cycle period. Each star is counted only once when the RV period is close to multiple activity-related periods (e.g., both rotation and magnetic cycle). This corresponds to a fraction of $28.2\%$ of stars with RV signals close to stellar rotation or activity periods.

\subsubsection{Stars with no planetary companion}
Among the 633 stars without known planetary companions, only 232 show significant period detections in the RV dataset. Table \ref{tab:RV_new} lists the stars for which RV periods are close to rotation or magnetic cycle periods reported in the literature. Additionally, we identify RV periods that are similar to those found in the S-index and/or BIS periodograms for stars lacking published rotation or magnetic cycle periods, to the best of our knowledge. However, some of these stars do not show a significant correlation between RVs and either of the activity indicators. This is the case, for instance, for HIP32103, HIP40133, and HIP85042 for short-period signals likely related to rotation, and for HIP25421, HIP54597, and HIP115577 for long-period signals possibly associated with magnetic cycles. These findings suggest that, despite the similarity in periods within uncertainties, the signals may not share a common physical origin. Some of these RV signals exhibit very large semi-amplitudes (e.g., HIP43297), suggesting that these periodicities may be caused by undetected planetary companions, even though their periods closely match those seen in the activity indicators. 50 out of 235 stars exhibit an RV period close to a rotation, magnetic cycle, S-index, or BIS period (including the three stars reassigned from the first subsample). Stars for which the RV period is close to both the rotation and magnetic cycle periods are counted only once. This corresponds to a fraction of $21.3\%$.
\begin{table*}
\begin{center}
\tiny
    \caption{\small Same as Table. \ref{tab:RV_comp} but for stars with no known planetary companion.}
    \vspace*{5mm}
    \label{tab:RV_new}
    \begin{tabular}{rrrrrrrr}
    \hline
HIP & Lit $P_{\rm{rot}}/P_{\rm{mag}}$ [d]& Ref. & $P_{\rm{S}}$ [d] & $P_{\rm{BIS}}$ [d] & $P_{\rm{RV}}$ [d] &$K_{\mathrm{RV}} [m s^{-1}]$& N\\
\hline
57 & --&-- &$16.9587 \pm 0.0024$& --& $16.89471 \pm 0.00090$ & $20.2 \pm 1.2$&37\\
1599 & $16.7 \pm 2.6$& \protect\cite{lovis2011harps} & $14.28872 \pm 0.00031 $& $14.6429 \pm 0.0020$ & $15.6555 \pm 0.0029$ & $0.645 \pm 0.083$& 3834\\
3979 & $21.3 \pm 3.2$ &\protect\cite{lovis2011harps} &$20.499 \pm 0.085  $&$18.65 \pm 0.084$ & $20.348 \pm 0.042$ &$3.94 \pm 0.61$& 60\\
5938 & --& --&$3.6897 \pm 0.0015$& $3.679758 \pm 0.000097$& $3.93169 \pm 0.00012$ &$32.5 \pm 1.6$& 30\\
12110 & --& --&$4.5836 \pm 0.0061$& $10.543 \pm 0.027$&  $5.11010 \pm 0.00036$ &$40.8 \pm 2.4$& 18\\
15442 & $22.3 \pm 3.2$ &\protect\cite{lovis2011harps}& $18.696 \pm 0.025$ & --& $19.199 \pm 0.075$ &$2.2 \pm 1.4$& 37 \\
20263 & $306.9 \pm 0.4$ &\protect\cite{auriere2008ek} &$304 \pm 31$ &$312 \pm 87$ & $296.15 \pm 0.43$ &$41.3 \pm 2.0$& 116\\
29525 & $7.80 \pm 1.0$& \protect\cite{pizzolato2003stellar}& --& $ 11.967 \pm     0.019$ &$8.83056 \pm 0.00024$ &$17.6 \pm 1.8$& 20\\
30260 & --& --&--& $14.18 \pm 0.18$& $14.102 \pm 0.023$ &$17.5 \pm 2.1$& 32\\
30979 & --& --&--& $14.629 \pm 0.053$& $13.4194 \pm 0.0046$ & $10.4 \pm 1.9$&19\\
32103 & --& --&--& $19.8224 \pm 0.0013$& $20.8665 \pm 0.0031$ &$14.9 \pm 1.4$& 35\\
37447 & --& --&$2.1493 \pm 0.0022$& $2.1683 \pm 0.0046$&  $2.28991 \pm 0.00084$ &$12.0 \pm 3.4$& 41\\
40133 & $28.0\pm 1.0$ &\protect\cite{wright2004chromospheric} & $ 25.82 \pm 0.83$ &--&$24.666 \pm 0.024$ &$5.4 \pm 1.5$& 27\\
41926 & $40.2\pm 4.1$ &\protect\cite{lovis2011harps}& $35.2408 \pm 0.0021$&--& $35.201 \pm 0.049 $&$0.58 \pm 0.23$ & 496\\
43297 & $17.0 \pm 1.0$ &\protect\cite{wright2004chromospheric} & $23.25 \pm 0.79$ & $17.109 \pm 0.014$ & $23.5529 \pm 0.0027$ &$84.4 \pm 3.5$& 27\\
43797 &--& --& $2.5050 \pm 0.0058$ & $2.32112 \pm 0.00014$& $2.13780 \pm 0.00018$ & $26.5 \pm 4.5$&44\\
44713 & $25.1 \pm 3.2$ &\protect\cite{lovis2011harps}& $43.0266 \pm 0.0092$ & $40.15231 \pm 0.00040$ & $22.1238 \pm 0.0068 $ &$2.27 \pm 0.45$& 148\\
46677 & $36.7 \pm 5.3$ &\protect\cite{lovis2011harps}  &$37.3460 \pm 0.0073$ &$32.641 \pm 0.033$&$18.5755 \pm 0.0069$ &$1.68 \pm 0.49$& 139\\
52316 & $58.7 \pm 4.9$ &\protect\cite{lovis2011harps}& --& --& $41.957 \pm 0.050 $&$3.87 \pm 0.70$ & 61\\
52369 & $18.6 \pm 2.9$ &\protect\cite{lovis2011harps} & $19.546 \pm 0.013$& $14.651 \pm 0.029$&$24.5509 \pm 0.0094$&$3.41 \pm 0.59$& 90\\
55210 & $33.5 \pm 3.8$ & \protect\cite{lovis2011harps} &$38.977 \pm 0.041$& $38.99\pm 0.39 $ & $39.418 \pm 0.021$ &$3.63 \pm 0.95$& 90\\
56445 & $7.7 \pm 1.0$  &\protect\cite{hempelmann2016measuring}& $ 11.06 \pm 0.32$ &-- & $9.6416 \pm 0.0032$ &$12 \pm 14$& 30\\  
61291 & $43.0 \pm 4.6$ &\protect\cite{lovis2011harps}  &$39.4608 \pm 0.0015 $&$48.152 \pm 0.010$ &$39.461 \pm 0.082$ &$0.58 \pm 0.18$& 839\\
70459 & $15.7 \pm 2.7$ &\protect\cite{lovis2011harps}& --& --& $36.971 \pm 0.027 $ &$5.89 \pm 0.66$& 75\\
79190 & $48.2 \pm 4.7$ &\protect\cite{lovis2011harps}  &$39.2574 \pm 0.0052$ &$38.609 \pm 0.027$ &$39.699 \pm 0.025$ &$1.47 \pm 0.43$& 438\\
79715 &-- &--& --& $53.20 \pm 0.15$& $50.766 \pm 0.036$ &$8.2 \pm 1.2$ &50\\
81819 & --&--& $45.486 \pm 0.026$& --& $44.34 \pm 0.10$ &$5.5 \pm 8.8$& 56\\
82588 & $11.4 \pm 1.0 $ &\protect\cite{pizzolato2003stellar} &--&$12.000 \pm 0.040$ &$8.7970 \pm 0.0020$&$7.9 \pm 1.8$ & 116\\
85042 & $31.0 \pm 3.4$ &\protect\cite{lovis2011harps}  &$32.2073 \pm 0.0012$&--&$38.682 \pm 0.024$&$2.65 \pm 0.67$ & 201\\
99174 &-- &--& $18.129 \pm 0.044$& --& $18.9610 \pm 0.0059$ &$18.9610 \pm 0.0059$& 25 \\
105184 &-- & --& $63.585 \pm 0.013$& $52.518 \pm 0.028$& $62.298 \pm 0.071$ &$3.57 \pm 0.47$& 175\\
108241 & --&--& $17.27 \pm 0.13$& $17.85 \pm 0.48$& $18.551 \pm 0.042$&$15.3 \pm 1.0$& 30 \\
114948 &-- &--& -- & $2.64712 \pm 0.00035$&$ 2.103104 \pm 0.000054$&$2.103104 \pm 0.000054 $&33 \\
\hline		 
3170 & --&-- & $1436.7 \pm 3.1$ & --& $1873 \pm 123$&$1.82 \pm 0.34$&256 \\
5280 & --&-- & $1599 \pm 40$ & --& $1590 \pm 37$ &$6.3 \pm 2.1$&33\\
9400 & --&-- & $2697 \pm 34$ &$ 2392 \pm 15$&$2878 \pm 222$ &$7.2 \pm 3.4$& 78 \\
11915 & --&-- & $2807 \pm 138$ & $3849 \pm 245$ &$3243 \pm 59$&$16.36 \pm 0.48$&96 \\ 
25421 & --&-- & $2354 \pm 15$ & -- &  $3138 \pm 490$&$2.5 \pm 1.5$ &74\\
34879 &-- &-- & $2352 \pm 161$ & -- & $2372 \pm 53$&$4.7 \pm 1.$ & 48\\
41926 & $3050 \pm 558$ &\protect\cite{lovis2011harps} & $3335.4 \pm 7.6$ & $3545.0 \pm 9.6 $&$4120 \pm 165$ &$3.93 \pm 0.44$& 496\\
42291 &$ 2801 \pm 984$ &\protect\cite{lovis2011harps} & $2703 \pm 121$ & --&$2262 \pm 33$ &$1.99 \pm 0.65$&227\\
46677 & $1534 \pm 37$ &\protect\cite{boro2018chromospheric} & $3772 \pm 29$ & $2356 \pm 254$ &$3435 \pm 62$ &$5.07 \pm 0.55$& 139\\
51297 & --&-- & $1527 \pm 437$ & -- & $2448 \pm 654$ &$3 \pm 11$& 28\\
54597 & --&-- & $2756 \pm 28$ & -- & $3361 \pm 25$ &$36.90 \pm 0.87$&61 \\
64408 & $897  \pm 61$ &\protect\cite{lovis2011harps}& $629 \pm 8.6$& $615.2 \pm 3.8$&$1200 \pm 22$ &$1.94 \pm 0.25$& 736\\
64408 & --& --& $1211 \pm 36$& $1142.3 \pm 2.7$& --&--&736\\
78170 &-- &-- & $3492 \pm 67$  & -- & $3531 \pm 500 $ &$6.6 \pm 6.9$& 51\\
79190 & $4103  \pm 1321$ &\protect\cite{lovis2011harps} & $5159 \pm 35$ & $2661.2 \pm 2.0$& $5403 \pm 1445$&$5.38 \pm 0.43$ & 438\\
111978 &-- &-- & $3475 \pm 170$ & -- & $3695 \pm 308$ &$10.0 \pm 9.5$& 64\\
115577 &-- &-- & $1824 \pm 35$ & -- &$1910 \pm 50 $ &$3.8 \pm 3.7$& 205\\
116937 & --&-- & $1848 \pm 457$ & $2531 \pm 5203$ & $1374 \pm 73$ &$27 \pm 111$& 30\\
\hline
\end{tabular}
\end{center}
\end{table*}

\subsection{Rotation and magnetic cycle periods}
\subsubsection{Rotation periods}
Nearly all stars in the sample with known planetary companions have estimated rotation periods reported in the literature. We provide updated estimates from the S-index and BIS in Table \ref{tab:Prot_comp}. For stars without planetary companions, Table \ref{tab:Prot_newLit} compares literature values with those derived from our analysis. Periods listed in Tables \ref{tab:RV_comp} and \ref{tab:RV_new} are not repeated here.

\begin{figure}
\begin{center}
\includegraphics[width=0.6\linewidth]{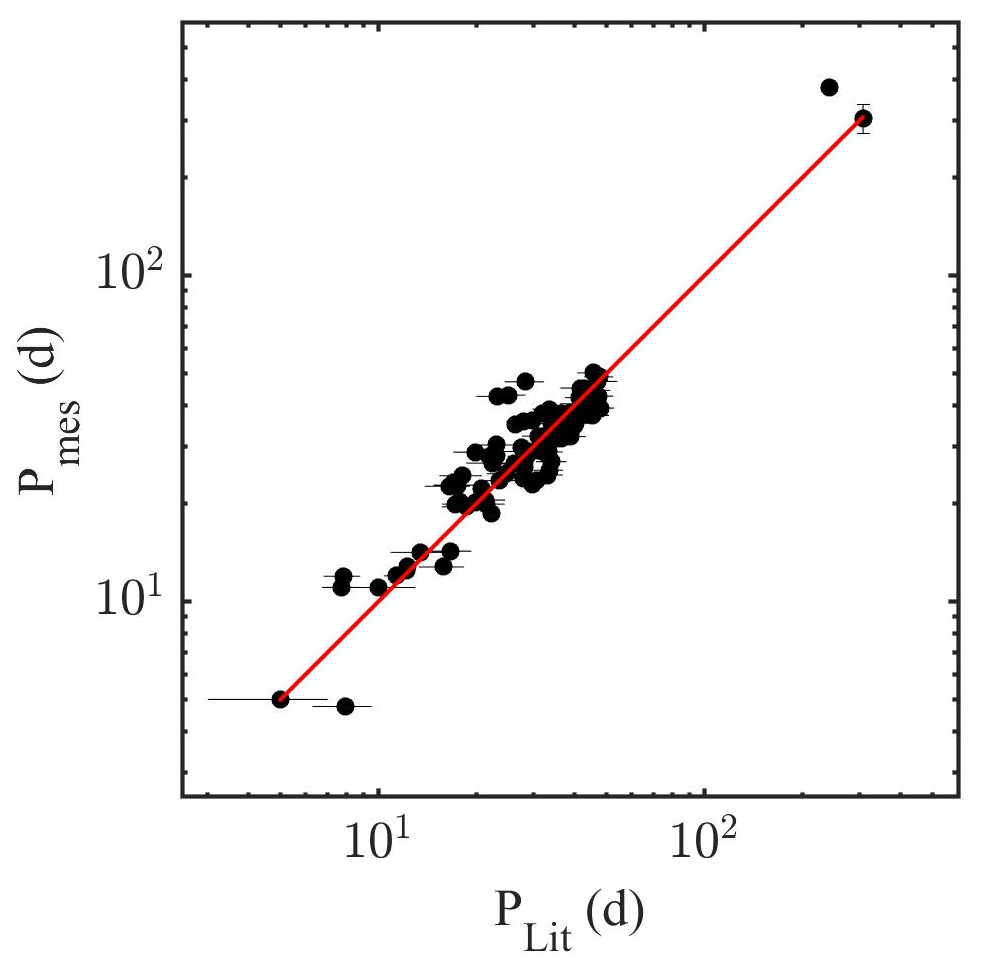}
\end{center}
\caption{Comparison between our measured rotation periods from S-index or BIS and those reported in the literature. The red line displays the 1:1 relation.}
\label{fig:PlitVSPmas}
\end{figure}

For most stars in our sample, the rotation periods derived from the S-index and BIS are broadly consistent with values reported in the literature, typically within the stated uncertainties. Figure~\ref{fig:PlitVSPmas} compares our derived rotation periods with literature values compiled for the full sample, including both stars with (Table. \ref{tab:RV_comp} and Table. \ref{tab:Prot_comp}) and without confirmed planetary companions (Table. \ref{tab:RV_new} and Table. \ref{tab:Prot_newLit}). No distinction is made between the methods used to derive them, whether photometric, spectroscopic, or based on activity–rotation relationships. We excluded stars for which only a harmonic (half the true period) was detected, as well as HIP14501, where a significantly different period has been suggested. Overall, the comparison shows a strong agreement between our results and previously published values.

\begin{figure}
\begin{subfigure}{.5\textwidth}
\includegraphics[width=\linewidth]{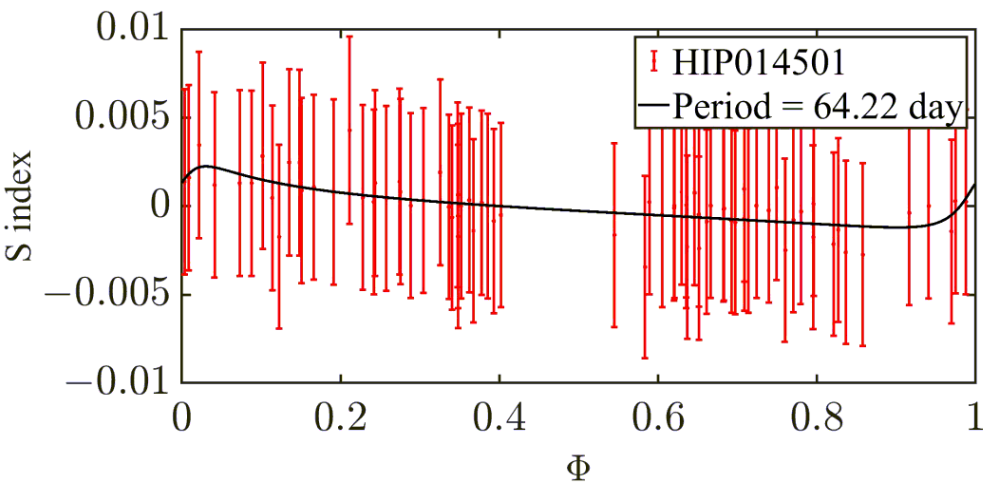}
\end{subfigure}
\begin{subfigure}{.5\textwidth}
\includegraphics[width=\linewidth]{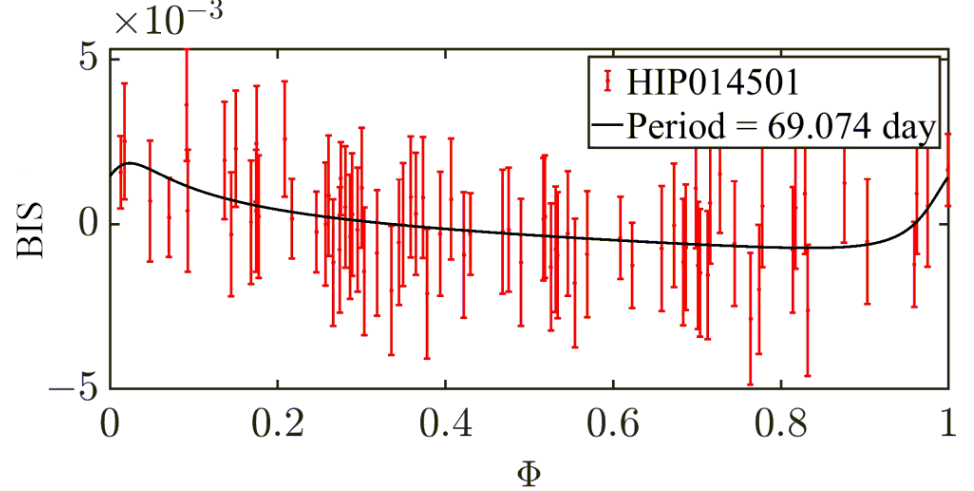}
\end{subfigure}
\caption{Phase-folded time-series of S-index data fitted with 64.2 d period (left panel) and BIS data fitted with 69.1 d period (right panel).}
\label{fig:HIP14501}
\end{figure}

However, several targets display significant discrepancies or complex behaviour that may reflect the underlying stellar activity dynamics. For example, HIP14501 shows much longer periods from both the S-index and BIS ($64.22 \pm 0.12$ d and $69.074 \pm 0.085$ d, respectively) compared to the literature ($24.9  \pm 2.5$ d \citep{crepp2014trends}), possibly indicating a previously misidentified rotation period. The phase-folded time series of S-index and BIS over these respective periods are provided in Fig. \ref{fig:HIP14501}. For HIP29432, although no rotation period derived directly from spectroscopic activity indicators has been reported in the literature, \citet{fulton2016three} identified a long-period signal of $\sim$4850~d in HIRES RVs, which they attributed to a magnetic activity cycle based on its presence in the S-index. Using \textit{CoRoT} photometry spanning 173~d, they also reported a broad peak around 16.9~d, interpreted as the stellar rotation period, although they noted that its exact value depends on the polynomial detrending applied to the light curve and could not be robustly constrained.
In our HARPS data, neither a short-period signal near 17~d nor a long-period magnetic cycle is detected in the S-index or BIS periodograms. Instead, after subtracting long-term trends, the residuals of both indicators reveal a consistent and well-defined period at 107.7~d. The strong agreement between the S-index and BIS at this timescale suggests that this signal is robust within our dataset and likely reflects a distinct activity-related timescale during the HARPS observing window. In HIP42401, the BIS period ($6.016331 \pm 0.000045$ d) is approximately half the S-index period ($12.44243 \pm 0.00025$ d), likely indicating a harmonic detection. Some stars (e.g., HIP38041, HIP52369, HIP60081) exhibit BIS periods significantly different from those of the S-index or literature, which may point to sensitivity to evolving surface features or differential rotation. Although both BIS and S-index are extracted from the same spectra, they probe different physical manifestations of stellar activity.
The S-index traces magnetic emission in the \ion{Ca}{ii} H\&K line cores, which is mainly produced by bright plage regions in the chromosphere \citep{sowmya2023modeling}, whereas the BIS is sensitive to line-profile distortions caused by photospheric magnetic structures \citep{dall2006bisectors}. Plages generally have longer lifetimes than spots, often persisting for several rotation cycles \citep{foukal2013solar}, and they are not always co-spatial with photospheric spots \citep{mandal2017association}. As a result, the S-index and BIS can be dominated by different active regions at a given time. Since the BIS response depends strongly on the projected position and latitude of photospheric structures \citep{boisse2011disentangling}, while the S-index is driven by chromospheric plages that may occupy different latitude bands, the two indicators can naturally exhibit different dominant periods. In contrast, stars like HIP74653 and HIP82632 exhibit S-index periods nearly half the literature values, possibly corresponding to harmonics of the underlying stellar rotation or due to evolving spot distributions or magnetic cycle-related modulation. Larger discrepancies are seen in stars such as HIP42291, HIP44713, and HIP74389, where the S-index period is considerably longer than the literature value, which may reflect spot migration or the ability of chromospheric indicators to trace longer-term activity variations which were missed previously due to lack of data.

For non-planet-hosting stars and no documented rotation period in the literature, we identify candidate rotation periods by comparing results from the S-index and BIS. When a short period is detected in only one of these indicators, we further verify it through a GLS period search in TESS photometric data obtained through the Mikulski Archive for Space Telescopes (MAST)\footnote{\url{https://mast.stsci.edu/portal/Mashup/Clients/Mast/Portal.html}}. We used data with an exposure time of 1800 seconds processed with TESS-SPOC pipeline. Periods detected in the S-index or BIS closely matching those found in RV measurements (Table. \ref{tab:RV_new})—and show a significant correlation between RV and either indicator—are also considered potential rotation periods (eg. HIP57 and HIP30260). Table \ref{tab:Prot_newUnknown} lists the newly proposed rotation periods for these stars, including only those detected in at least two independent datasets (S-index, BIS, or photometry). In some cases, only half of the period is detected in one of the datasets (e.g., HIP80, HIP12110, and HIP47681). It is important to note that periods identified in photometric data may not exactly match those in the S-index or BIS. In addition to the effects of differential rotation, such differences can arise because photometric and spectroscopic observations are often not obtained simultaneously, allowing active regions to evolve or migrate between observing epochs. HIP105184 shows two dominant periods in the S-index ($23.0430 \pm 0.0013$ d and $63.585 \pm 0.013$ d). Since the BIS displays a similar $26.894020 \pm 0.000013$ d signal and the RV exhibits a $62.298 \pm 0.071$ d signal, these periodicities are likely aliases of the true stellar rotation period. We are cautious about the proposed rotation period of HIP 8119, as it is based on only 12 measurements.

\subsubsection{Magnetic cycle periods}
The derivation of an empirical relation between magnetic cycle periods and stellar rotation periods is hindered by the limited number of well-characterised activity cycles. To address this limitation, long-term time-series observations are essential. In this work, we take advantage of recent HARPS data to provide new estimates of magnetic cycle periods.

For stars with confirmed planetary companions, we compare long periods detected in the S-index and BIS with magnetic cycle periods reported in the literature (Table \ref{tab:Pmag_compLit}). For those without known literature values, possible cycle periods inferred from S-index and BIS data are listed in Table \ref{tab:Pmag_compUnknown}, where any long period is treated as a candidate magnetic cycle. For stars without detected planetary companions, Table \ref{tab:Pmag_newLit} shows a comparison between literature magnetic cycle periods and those derived from our analysis, while Table \ref{tab:Pmag_newUnknown} lists stars with no known values, for which we propose new cycle periods based on S-index and/or BIS data. Periods listed in Tables \ref{tab:RV_comp} and \ref{tab:RV_new} are not repeated here.

For many stars, especially those with longer magnetic cycles, the periods derived from the S-index and BIS data are broadly consistent with each other and often agree with values reported in the literature. However, significant mismatches also occur in some cases. For instance, in HIP15510, the BIS suggests a much longer cycle than reported in the literature, while in HIP33229, the S-index indicates a period nearly twice as long as the literature value. For stars with short cycle periods (e.g., $<1000$ d), detections can be affected by aliasing, sparse time sampling, or contamination from rotational modulation. In contrast, long-period solutions tend to have larger uncertainties or are absent from the literature, likely due to insufficient observational baselines. Stars such as HIP16085 and HIP27887 show long cycles that are only now becoming detectable thanks to extended monitoring efforts.

Some cases, like HIP46677, show significant disagreement among all three sources: the literature reports the shortest period, the BIS gives an intermediate value, and the S-index yields the longest cycle. This may point to evolving magnetic behaviour or misidentification of periodic signals. Discrepancies between BIS and S-index estimates in several stars likely arise from their differing sensitivities to surface activity features, such as spots versus plages.

In addition to the primary magnetic cycle periods, several stars in our sample exhibit evidence of secondary long-term periodicities. These include HIP436, HIP1599, HIP10301, HIP30503, HIP40693, HIP48331, HIP64408, and HIP79190.
For example, HIP64408 shows two periodicities at $629 \pm 8.6$ d and $1211 \pm 36$ d in the S-index, which could indicate a 2:1 harmonic ratio or modulation in the activity pattern. HIP79190 exhibits a notably shorter cycle in the BIS compared to its S-index and literature values, possibly suggesting a distinct variability mechanism. In some stars, such as HIP1599, the presence of multiple peaks may also be due to aliasing effects.

The presence of secondary magnetic cycles in our sample is reminiscent of solar behaviour. The Sun exhibits not only the well-known 11-year sunspot cycle, but also longer-term variations like the ~80-year Gleissberg cycle, as well as hemispheric asymmetries that produce other periodicities, e.g., 51.34, 8.83, and 3.77 years \citep{deng2016systematic}. The detection of multiple cycles in stars with solar-like or longer rotation periods aligns with this picture. Interestingly, while \citet{boro2018chromospheric} reported secondary cycles predominantly in fast rotators, our findings show such behaviour in stars with rotation periods similar to or longer than the Sun’s.

\section{Discussion}
\label{sect:discuss}
\subsection{RV Signals and Stellar Activity}
Stars with confirmed planets show a higher rate of RV signals close to rotation/activity periods (28.2\%) than stars without planets (21.3\%), but this difference is not statistically significant since it has a p-value of 0.14. The trend might be real, but with our current sample size, there is a 14\% chance that this difference arises purely by chance. This dataset alone is insufficient to conclude that stars with planets are more prone to RV confusion with rotation or magnetic cycle periods than those without.

To compare the distribution of RV periods coinciding with stellar rotation/activity cycle periods between planet-hosting and non-planet-hosting stars, we applied two complementary non-parametric statistical tests. The two-sample Kolmogorov–Smirnov (KS) test probes differences in the overall shapes of the distributions, making it sensitive to changes in spread or tails, but it is relatively insensitive to shifts in central tendency for modest sample sizes. The Mann–Whitney U test, on the other hand, is particularly sensitive to systematic shifts in central tendency (median values), but does not capture differences in distribution shape. Using both tests therefore allows us to assess whether the two samples differ either globally or through a systematic offset in typical RV–activity period proximity. The KS test yields a statistic of 0.13 and a p-value of 0.79, while the Mann–Whitney U test results in a p-value of 0.87, indicating no significant difference between the two samples. This suggests that the proximity of RV periods to rotational/ activity cycle periods occurs across similar timescales regardless of planetary status.

\subsection{Rotation versus Magnetic Cycles}
\subsubsection{Activity branches} 
In efforts to understand the stellar dynamo, earlier studies by \cite{brandenburg1998time} and \cite{saar1999time} identified two distinct branches—termed active and inactive—when plotting the ratio of the rotation period to the activity cycle period against the inverse Rossby number ($Ro^{-1}$). These two branches are separated by the so-called Vaughan-Preston gap \citep{vaughan1980survey}. For each branch, they derived the following linear relations of the form: $\log P_{\mathrm{rot}}/P_{\mathrm{mag}} = -a\,\log P_{\mathrm{rot}} + b$: for the active branch, $a_A = 0.458, \, b_A = -2.355$; and for the inactive branch, $a_I = 0.475, \, b_I = -3.077$. 

Recent studies have yielded contrasting results. \cite{olah2016magnetic} reported a relation of $P_{\mathrm{mag}} \propto P_{\mathrm{rot}}^{0.24}$, suggesting a negative slope. \cite{boro2018chromospheric} and \cite{olspert2018estimating} questioned the presence of the two distinct branches with a positive slope.  In contrast, both a previous study by \citet{brandenburg2017evolution} and a more recent one by \citet{mittag2023revisiting} supported the existence of such a positive trend.

\begin{figure}
\begin{center}
\includegraphics[width=0.8\linewidth]{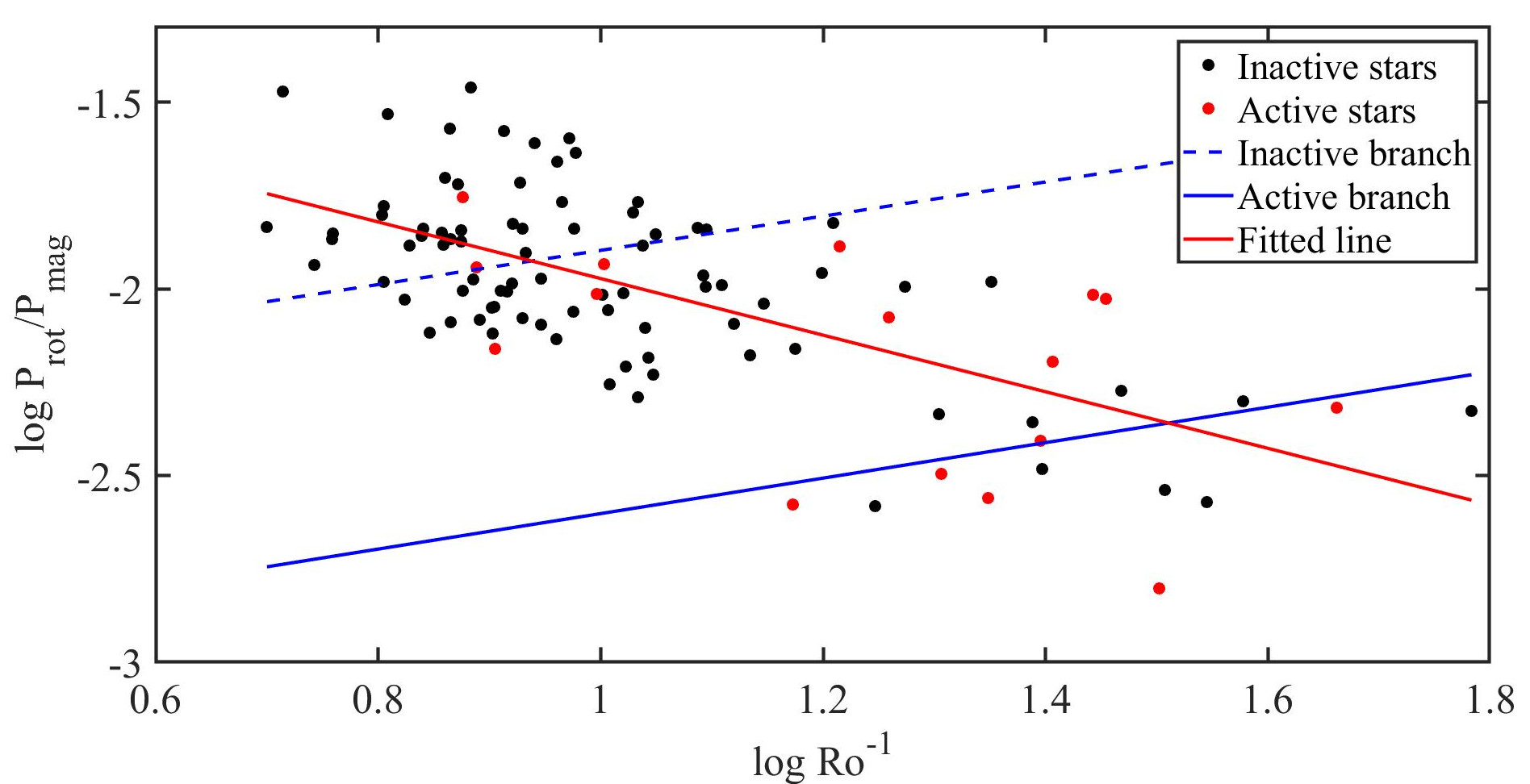}
\end{center}
\caption{$P_{\rm{rot}}/P_{\rm{mag}}$ vs. $Ro^{-1}$ in logarithmic scale, where active stars are shown in red and inactive stars are in black. Active and inactive branches from \protect\cite{saar1999time} are shown as blue solid and dashed lines, respectively. Our fit is shown as a red solid line.}
\label{fig:ProtPmag_Ross}
\end{figure}

To test the presence of the active and inactive branches, we examine the relationship between stellar rotation and magnetic cycle periods using both the values derived in this work and those compiled from the literature. Following previous studies, we plot the ratio $P_{\mathrm{rot}}/P_{\mathrm{mag}}$ against the inverse Rossby number $Ro^{-1}$ on a logarithmic scale, as shown in Fig. \ref{fig:ProtPmag_Ross}. The Rossby number is defined as $Ro = 1/2\Omega\tau_c$, where $\Omega$ is the stellar rotation frequency and $\tau_c$ is the convective turnover time. We compute $\tau_c$ using the empirical relation provided by \citet{noyes1984rotation}. For consistency across the sample, when a target exhibits multiple magnetic cycles, we adopt the longer one. If both the rotation and magnetic cycle periods are available from our S-index analysis, we use those values; otherwise, we fill in any missing values with data from the literature. Since the uncertainties for different periods are heterogeneous and not directly comparable across the compiled datasets; particularly those in our measurements are significantly smaller than those reported in previous studies, we apply unweighted linear least-squares fit in log–log space to characterize the global trend, to avoid biasing the results toward our more precise estimates. 

Including both stars with and without planetary companions, we observe in Fig. \ref{fig:ProtPmag_Ross} a general trend with a negative slope. The stars in our sample are spread across both the active and inactive branches, with no clear separation between them. Moreover, there is no distinct clustering between active stars with $\rm{log}R_{\rm{HK}}> -4.75$ (shown in red) and inactive stars with $\rm{log}R_{\rm{HK}}\leq -4.75$ (shown in black). We fit this trend with a slope of $a = -0.758\pm 0.044 $, which is consistent with the value reported by \cite{olah2016magnetic}, $-0.76\pm 0.15$. Although \cite{boro2018chromospheric} reported a similar negative slope, they argued that it is likely non-physical and attributed it to selection biases in the sample. Furthermore, \cite{warnecke2018dynamo}, using three-dimensional magnetohydrodynamical simulations of global convective dynamo models for solar-like stars, found no evidence for the distinct activity branches proposed in earlier observational studies. Instead, their results exhibited a negative slope consistent with the so-called transitional branch suggested by \cite{distefano2017activity} and further supported by \cite{boro2018chromospheric} and \cite{olspert2018estimating}. 

The observed negative slope of $P_{\mathrm{rot}}/P_{\mathrm{mag}}$ versus the inverse Rossby number $Ro^{-1}$ is also in agreement with the earlier study by \cite{baliunas1996dynamo}, which showed that the ratio $P_{\mathrm{rot}}/P_{\mathrm{mag}} $ is proportional to $ D^i$, where the dynamo number $D$ is proportional to $\Omega$ for solar and stellar dynamos. The exponent $i$ is typically $\geq 1/3$, depending on the specific dynamo regime. This trend reflects the expectation that the ratio $P_{\mathrm{rot}}/P_{\mathrm{mag}}$ increases with rotation period,  as turbulent effects ($\alpha$-effect), which are key to dynamo efficiency, are enhanced in the slow rotation regime and scale with $\Omega$ \citep{olspert2018estimating}. Correspondingly, the positive correlation between magnetic cycle and rotation periods across our sample suggests that stars with slower rotation rates tend to have longer magnetic cycles. This trend is consistent with dynamo theory, which posits that slower rotators experience reduced rotational shear ($\alpha$), diminishing the efficiency of poloidal field generation and lengthening the magnetic cycle period. Similar conclusions were reached by \citet{vashishth2023modelling} and \citet{hazra2019exploring}, who showed that increasing the rotation rate in slow rotators leads to shorter magnetic cycle periods

\begin{figure}
\begin{center}
\includegraphics[width=0.8\linewidth]{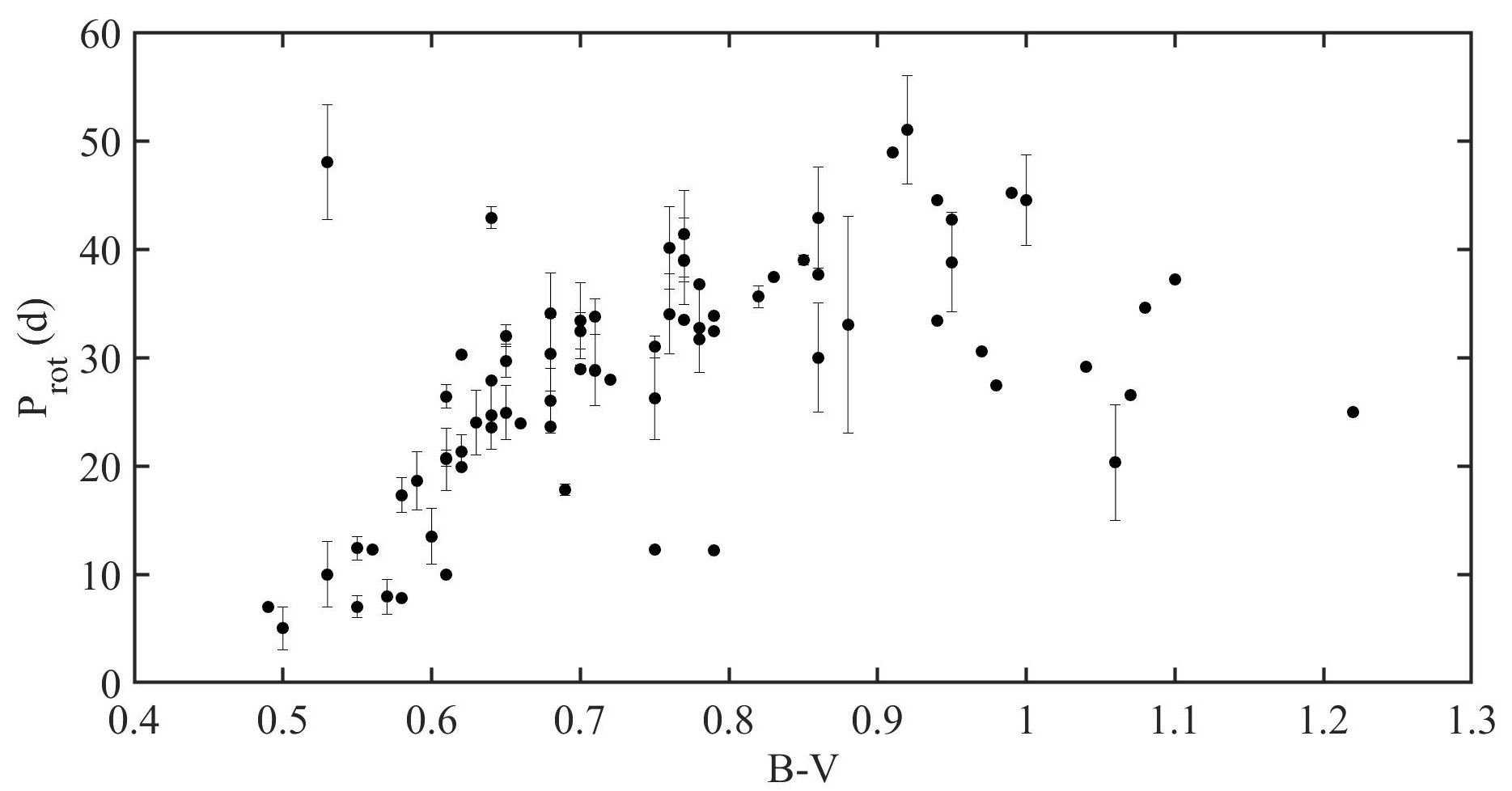}
\end{center}
\caption{Rotation period as a function of the colour index $B-V$ for stars with confirmed planetary companions.}
\label{fig:ProtB_V}
\end{figure}

In contrast to \cite{mittag2023revisiting}, we did not observe a dependence of the $P_{\rm{mag}}$–$Ro$ relation on the colour index $B-V$ (extracted from \cite{boro2018chromospheric}). However, a significant correlation is found between the rotation period and the colour index $B-V$ for stars with known planetary companions, as shown in Fig. \ref{fig:ProtB_V}, with a correlation coefficient of 0.54 and a p-value of zero. This correlation is absent for stars without known companions.

\subsubsection{Planetary effect} 
\begin{figure}
\begin{center}
\includegraphics[width=0.8\linewidth]{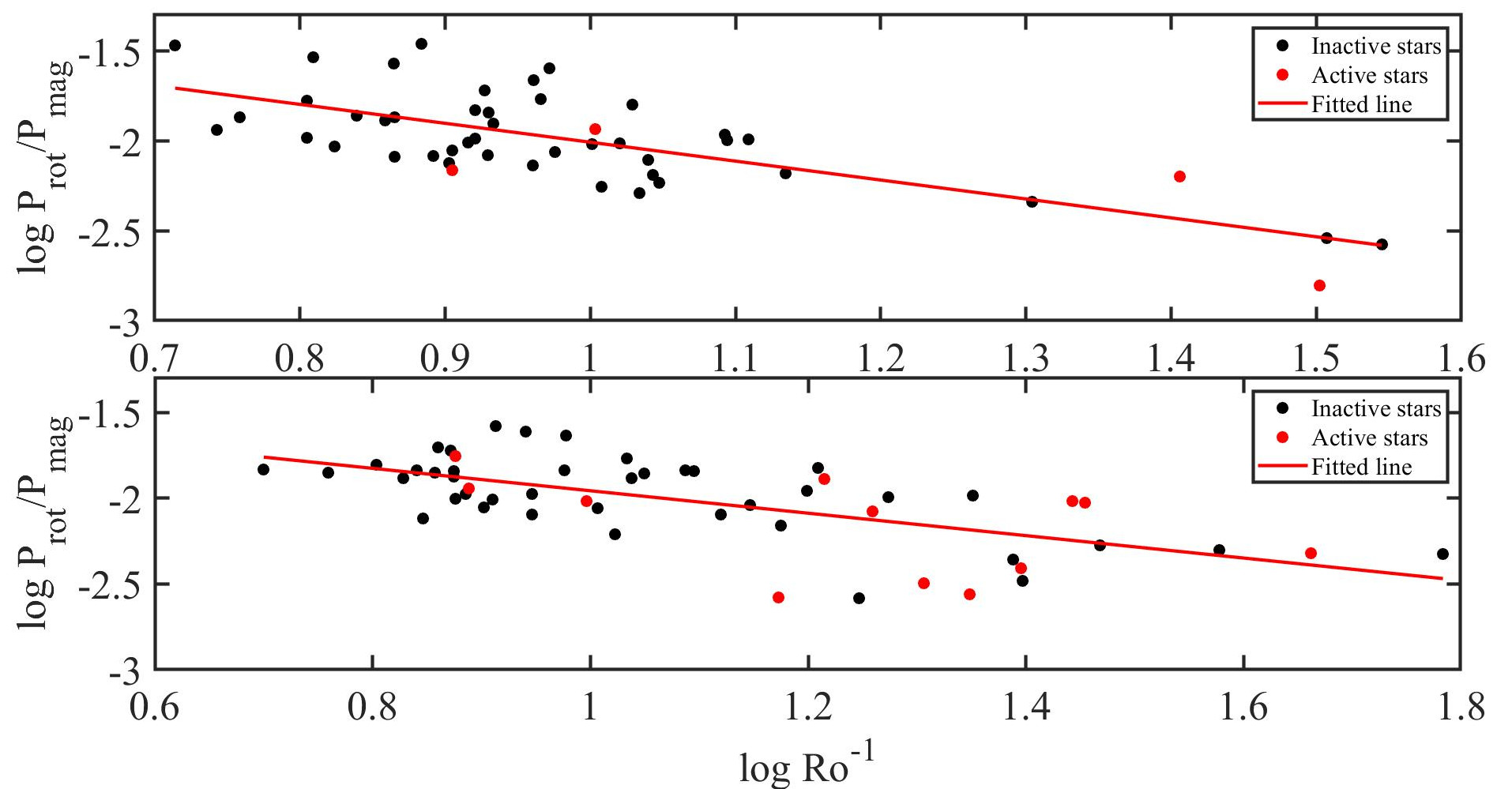}
\end{center}
\caption{$P_{\rm{rot}}/P_{\rm{mag}}$ vs. $Ro^{-1}$ in log scale, where active stars are shown in red and inactive are in black. Top panel: planet-hosting stars. Bottom panel: non-planet-hosting stars. Our fit is shown as a red solid line.}
\label{fig:ProtPmag_Ross_CompNew}
\end{figure}

To explore whether the relationship between magnetic cycle and rotation periods depends on the presence of planetary companions, we show $P_{\mathrm{rot}}/P_{\mathrm{mag}}$ versus $Ro^{-1}$ in Fig. \ref{fig:ProtPmag_Ross_CompNew}: the top panel displays stars with confirmed companions, while the bottom panel shows stars without confirmed companions.

Most stars with confirmed planetary companions in our sample are inactive ($\log R_{\mathrm{HK}} \leq -4.75$), with only four exceptions (HIP 801, HIP 12653, HIP 26013, HIP 70170). When examining the relation between the magnetic cycle and rotation periods for this subset, we find a slope of $-1.049 \pm 0.078$, suggesting no significant correlation. This lack of correlation may indicate that the presence of planetary companions disrupts or modifies the typical magnetic cycle–rotation relationship.

In contrast, stars without known planetary companions show a well-defined trend, with a slope of $-0.654 \pm 0.056$. This reveals a clear correlation between magnetic cycle and rotation periods, supporting the idea of a shared dynamo behaviour within this subsample. However, this slope is lower than those reported in previous studies, such as $-0.74$ by \citet{baliunas1996dynamo}, $-0.81 \pm 0.05$ by \citep{olah2009multiple}, and $-0.76\pm 0.15$ by \cite{olah2016magnetic}. The discrepancy may be explained by the fact that those earlier samples likely included a mix of stars with and without planets. Supporting this, when we combine all stars in our dataset—regardless of planetary companions—the slope becomes $-0.758 \pm 0.044$, closely matching the literature values.

The observed difference in behaviour between stars with and without planets points toward a possible influence of planetary companions on the stellar dynamo. Several studies have explored this hypothesis. For instance, \cite{bolmont2012effect} and \cite{ceillier2016rotation} argued that only close-in giant planets can affect the host star's rotation. \cite{alves2010rotational} found that stars without planets generally exhibit lower angular momentum than those with planets, especially at higher stellar masses. \cite{gonzalez2015parent} reported that stars that host planets, regardless of the type of planet, tend to rotate more slowly, with the slowest rotation observed in hosts of giant planets. \cite{wright2015magnetism} and others \citep{cohen2009interactions,lanza2008hot,lanza2009stellar} proposed that Hot Jupiters may influence stellar activity and magnetic behaviour through tidal or magnetic interactions. Furthermore, \cite{poppenhaeger2014indications} showed that in binary systems with strong tidal interactions, planet-hosting stars exhibit higher magnetic activity than their planet-free companions, suggesting that Hot Jupiters may inhibit stellar spin-down via angular momentum transfer or early star–disk interactions.

\begin{figure}
\begin{center}
\includegraphics[width=0.8\linewidth]{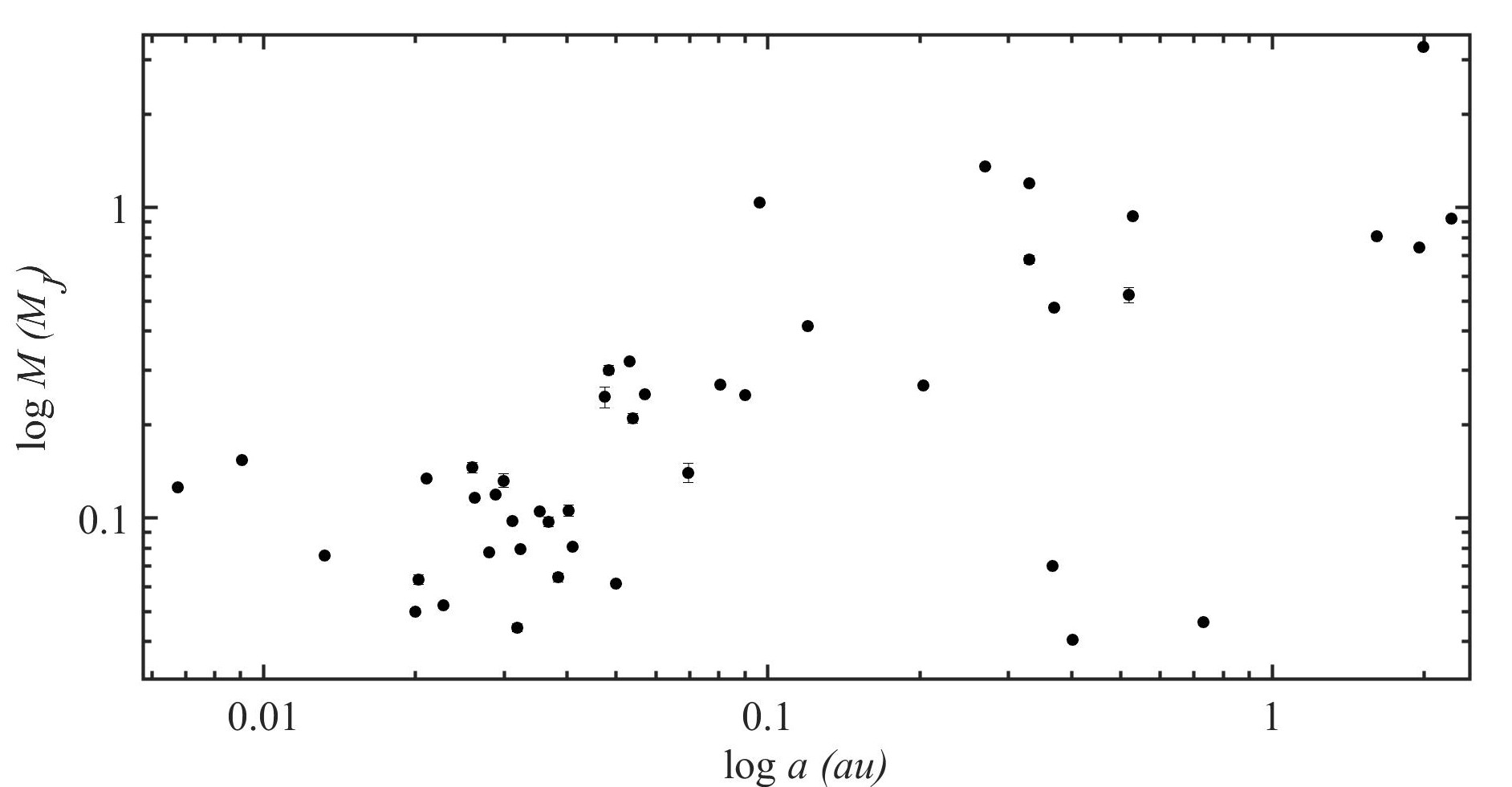}
\end{center}
\caption{Planetary mass in $M_{J}$ versus the semi-major axis in AU on a logarithmic scale.}
\label{fig:massVSa}
\end{figure}

In our sample, where we consider only the most massive and closest planet in multi-planet systems, the planets are predominantly close-in, with semi-major axes ranging from 0.041 to 3.30 AU with a median of $0.134 \pm 0.064$ AU, and masses from 0.0068 to 2.27 $M_J$ with a median of $0.0484 \pm 0.035$ $M_J$, as illustrated in the logarithmic plot of planetary mass versus semi-major axis (Figure~\ref{fig:massVSa}). These findings support the idea that even low-mass, close-in planets may play a role in shaping the host star’s magnetic dynamo.

Altogether, these results support the hypothesis that planetary companions, particularly close-in ones, can alter the magnetic and rotational evolution of their host stars. This may explain the differing magnetic cycle–rotation relationships observed in our sample and emphasises the need to consider planetary systems when interpreting stellar activity trends.

\section{Conclusions}
\label{conc}
In this work, we analysed RV, S-index, and BIS data for 767 stars from the catalogue of Gaia RV standard stars to identify periodicities associated with stellar rotation and magnetic activity cycles. We find that 28.2\% of stars with confirmed planetary companions exhibit RV signals close to rotation periods, magnetic cycle periods, or those detected in at least one activity indicator, compared to 21.3\% among stars without known companions. Although this suggests a slightly higher likelihood of confusion between RV and activity signals in planet-hosting stars, statistical tests (Kolmogorov–Smirnov and Mann–Whitney U) show that this difference is not statistically significant. As a result of our analysis, we identify three RV signals (HIP7240, HIP28460, and HIP48331) that are likely caused by stellar activity rather than planetary companions. We also flag several additional systems as requiring further investigation. In total, we provide new measurements of rotation periods for 125 stars (including 30 newly estimated) and detect magnetic cycle periods in 127 stars, 95 of which are reported for the first time. While there is broad agreement between rotation and cycle periods derived from S-index and BIS measurements, discrepancies in some cases suggest differences in the physical regions probed by each indicator or the presence of multiple activity cycles, similar to what is observed in the Sun. In particular, we report evidence for secondary cycles in several stars (HIP10301, HIP30503, HIP40693, HIP48331, and HIP64408).

We further examined the relationship between rotation and magnetic cycle periods within the context of stellar dynamo theory. Most stars with planets appear magnetically inactive, yet the full sample does not show a clear separation into active and inactive dynamo branches in the $P_{\rm rot}/P_{\rm cyc}$–Rossby number plot.  Instead, the data follow a global trend with a negative slope, consistent with recent dynamo models and the so-called transitional branch. The absence of a clear separation may reflect the effects of differential rotation or intrinsic variability in stellar dynamos, challenging previous studies that identified distinct dynamo branches. Notably, stars without planets show a tighter correlation, whereas planet-hosting stars exhibit a more scattered and steeper trend. This difference may indicate that planetary companions—especially close-in ones—can alter the host star's dynamo.

Our results highlight the importance of combining multiple activity diagnostics when interpreting RV signals and suggest that the traditional picture of discrete dynamo branches may not fully describe the diversity of magnetic behaviour in solar-type stars. Expanding the sample and improving the precision of rotation and cycle period measurements will be crucial for further testing these trends.

\begin{acknowledgements}
This research was achieved using the POLLUX database ( \url{http://pollux.oreme.org} ) operated at LUPM  (Université Montpellier - CNRS, France) with the support of the PNPS and INSU. This paper includes data collected with the TESS mission, obtained from the MAST data archive at the Space Telescope Science Institute (STScI). Funding for the TESS mission is provided by the NASA Explorer Program. STScI is operated by the Association of Universities for Research in Astronomy, Inc., under NASA contract NAS 5–26555. We thank Prof. Tiago M. D. Pereira for his helpful comments and suggestions. We thank the anonymous reviewer for providing constructive suggestions that improved the clarity of the paper.

Based on observations collected at the European Organisation for Astronomical Research in the Southern Hemisphere under ESO programme(s) 072.C-0488(E), 183.C-0972(A), 096.C-0499(A), 086.C-0284(A), 190.C-0027(A), 082.C-0212(A), 085.C-0063(A), 60.A-9709(G), 074.C-0364(A), 192.C-0852(A), 196.C-1006(A), 60.A-9036(A), 60.A-9109(A), 085.C-0019(A), 0100.C-0097(A), 0101.C-0379(A), 0102.C-0558(A), 0103.C-0432(A), 087.C-0831(A), 089.C-0732(A), 090.C-0421(A), 091.C-0034(A), 092.C-0721(A), 093.C-0409(A), 095.C-0551(A), 099.C-0458(A), 096.C-0460(A), 098.C-0366(A), 198.C-0836(A), 091.C-0936(A), 0103.C-0206(A), 0102.C-0584(A), 073.C-0784(B), 074.C-0012(A), 076.C-0878(A), 077.C-0530(A), 078.C-0833(A), 079.C-0681(A), 083.C-1001(A), 084.C-0229(A), 085.C-0318(A), 086.C-0230(A), 089.C-0050(A), 090.C-0849(A), 092.C-0579(A), 093.C-0062(A), 094.C-0797(A), 095.C-0040(A), 096.C-0053(A), 097.C-0021(A), 072.C-0513(B), 082.C-0312(A), 080.D-0047(A), 080.D-0408(A), 081.D-0066(A), 082.D-0499(A), 083.D-0549(B), 084.D-0591(C), 086.D-0078(D), 091.D-0469(A), 098.C-0739(A), 192.C-0224(A), 188.C-0265(A), 188.C-0265(D), 188.C-0265(F), 188.C-0265(G), 188.C-0265(H), 188.C-0265(I), 188.C-0265(K), 188.C-0265(L), 188.C-0265(M), 188.C-0265(N), 188.C-0265(O), 188.C-0265(P), 188.C-0265(R), 0100.D-0444(A), 0104.C-0090(A), 078.C-0044(A), 192.C-0852(M), 1102.C-0923(A), 0101.C-0275(B), 0104.C-0090(B), 188.C-0265(J), 079.C-0463(A), 074.C-0102(A), 078.C-0751(A), 079.C-0657(C), 081.C-0802(C), 082.C-0427(A), 074.C-0012(B), 078.D-0067(A), 091.C-0853(A), 60.A-9700(G), 076.C-0155(A), 074.C-0037(A), 192.C-0224(G), 087.C-0990(A), 088.C-0011(A), 0102.C-0451(B), 0102.C-0525(A), 0102.D-0483(A), 0101.C-0510(D), 077.C-0080(A), 078.C-0233(A), 078.C-0233(B), 086.D-0647(B), 0101.A-9016(A), 0100.C-0414(A), 0100.C-0414(B), 0101.C-0232(A), 0102.C-0338(C), 0103.A-9011(A), 0104.A-9004(A), 0100.A-9006(A), 086.D-0082(A), 088.C-0879(A), 097.C-0090(A), 099.A-9009(A), 1102.C-0249(A), 0102.A-9006(A), 0101.C-0829(A), 1102.C-0923(C), 081.C-0148(A), 081.C-0148(B), 088.C-0662(A), 088.C-0662(B), 089.C-0497(B), 091.C-0866(C), 0100.C-0487(A), 082.C-0212(B), 072.C-0513(D), 076.C-0878(B), 185.D-0056(B), 185.D-0056(H), 188.C-0265(B), 188.C-0265(E), 0100.C-0474(A), 0103.D-0445(A), 094.C-0901(A), 0101.C-0232(C), 0102.C-0338(A), 0102.C-0338(B), 188.C-0265(Q), 188.C-0265(C), 076.C-0279(A), 076.C-0279(B), 076.C-0279(C), 078.C-0209(A), 078.C-0209(B), 080.C-0664(A), 080.C-0712(A), 081.C-0774(A), 184.C-0815(A), 184.C-0815(E), 184.C-0815(F), 078.C-0751(B), 081.C-0802(B), 082.C-0427(C), 086.C-0145(A), 089.C-0415(B), 089.C-0497(A), 077.C-0295(A), 077.C-0295(B), 077.C-0295(D), 184.C-0815(C), 0103.C-0548(A), 079.C-0927(C), 087.C-0368(A), 091.C-0866(A), 091.C-0866(B), 075.C-0332(A), 183.D-0729(A), 073.D-0578(A), 082.C-0315(A), 083.C-0794(C), 083.C-0794(D), 077.C-0101(A), 083.C-0794(A), 077.C-0364(E), 093.C-0919(A), 192.C-0852(H), 081.C-0388(A), 084.C-0228(C), 087.C-0368(B), 072.C-0096(E), 073.D-0038(B), 073.D-0038(C), 073.D-0038(D), 074.D-0131(E), 075.D-0194(A), 076.D-0130(E), 078.D-0071(E), 079.D-0075(A), 079.D-0075(B), 079.D-0075(C), 079.D-0075(E), 080.D-0086(E), 081.D-0065(A), 081.D-0065(B), 081.D-0065(C), 081.D-0065(D), 081.D-0065(E), 072.C-0096(D), 079.C-0927(B), 081.C-0842(B), 192.C-0852(G), 108.22KV.001, 075.C-0202(A), 196.C-0042(D), 095.D-0026(A), 097.D-0150(A), 105.2045.001, 105.2045.002, 105.20L8.002, 190.D-0237(E), 075.C-0689(A), 075.C-0689(B), 077.C-0295(C), 081.C-0842(D), 073.D-0590(A), 0101.C-0275(A), 196.C-0042(A), 183.D-0729(B), 196.C-0042(E), 0101.C-0232(B), 097.C-0571(A), 099.C-0205(A), 087.C-0012(B), 0101.D-0494(A), 192.C-0224(B), 192.C-0224(C), 192.C-0224(H), 184.C-0815(B), 099.C-0491(A). 
\end{acknowledgements}

\appendix  
\section{Activity periods in RV dataset}
\begin{figure}
\begin{subfigure}{.5\textwidth}
\includegraphics[width=\linewidth]{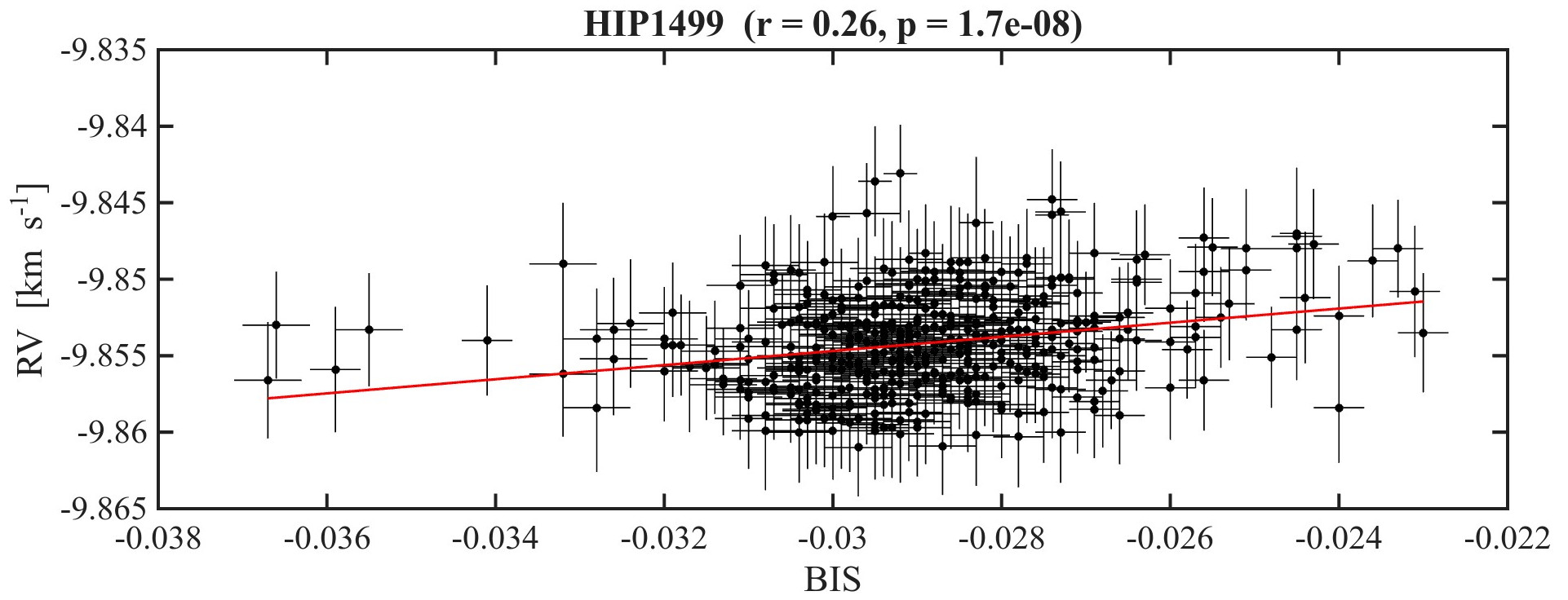}
\end{subfigure}
\begin{subfigure}{.5\textwidth}
\includegraphics[width=\linewidth]{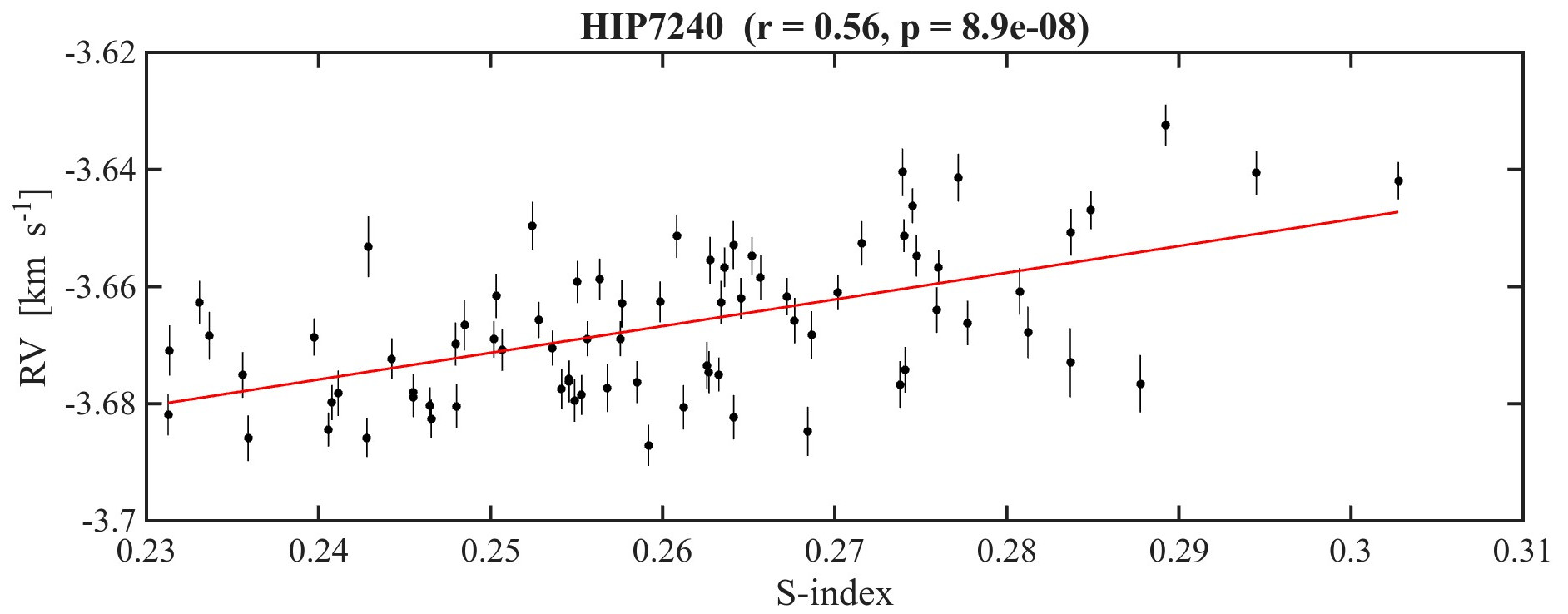}
\end{subfigure}
\begin{subfigure}{.5\textwidth}
\includegraphics[width=\linewidth]{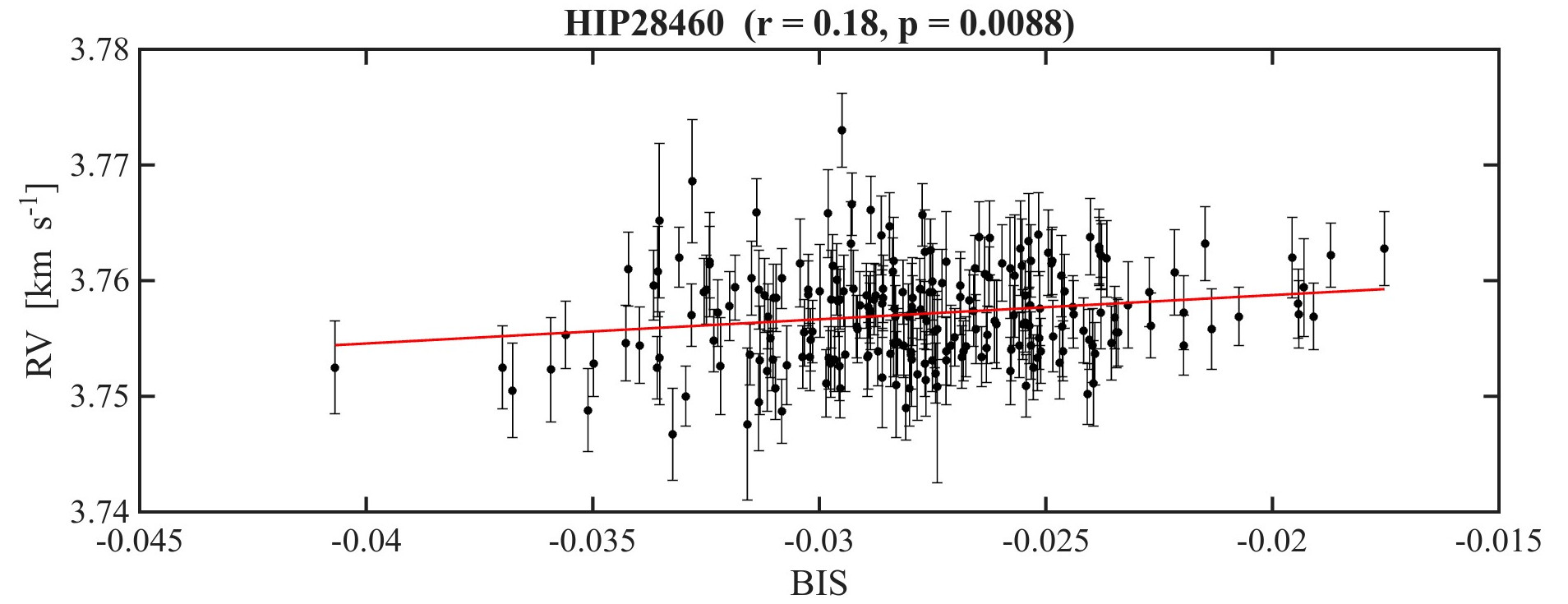}
\end{subfigure}
\begin{subfigure}{.5\textwidth}
\includegraphics[width=\linewidth]{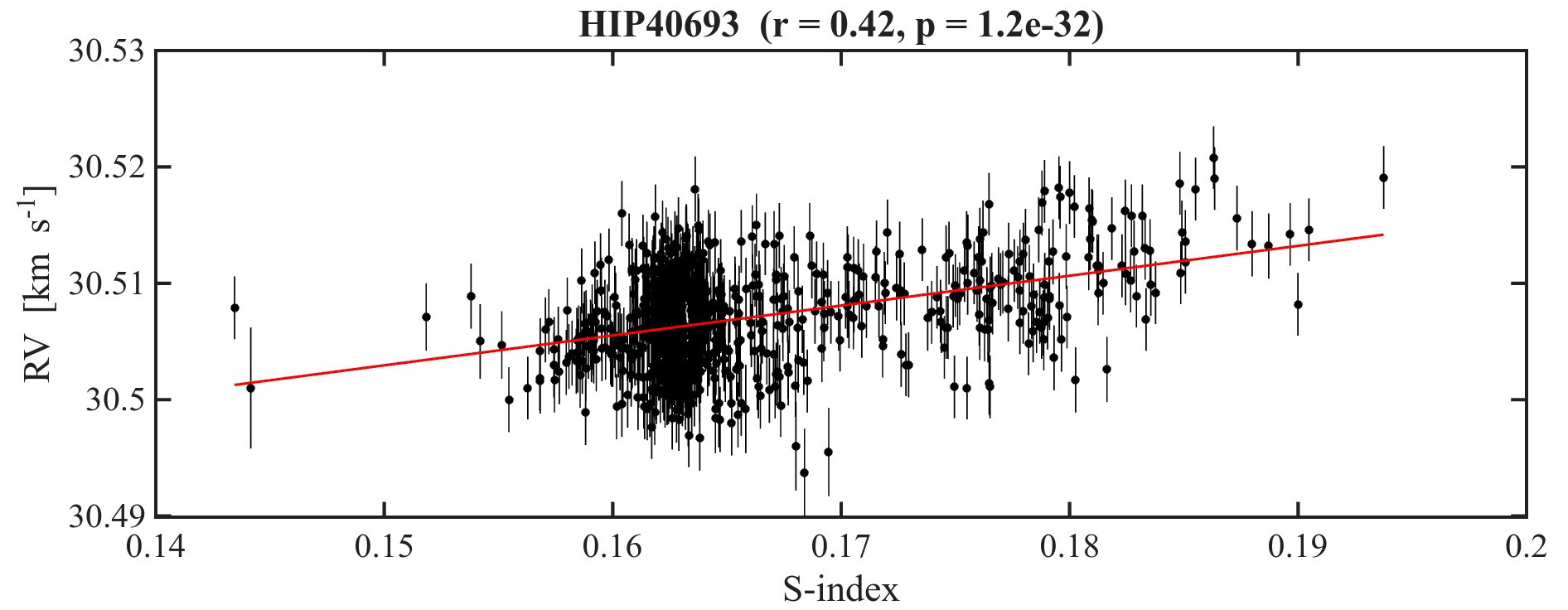}
\end{subfigure}
\begin{subfigure}{.5\textwidth}
\includegraphics[width=\linewidth]{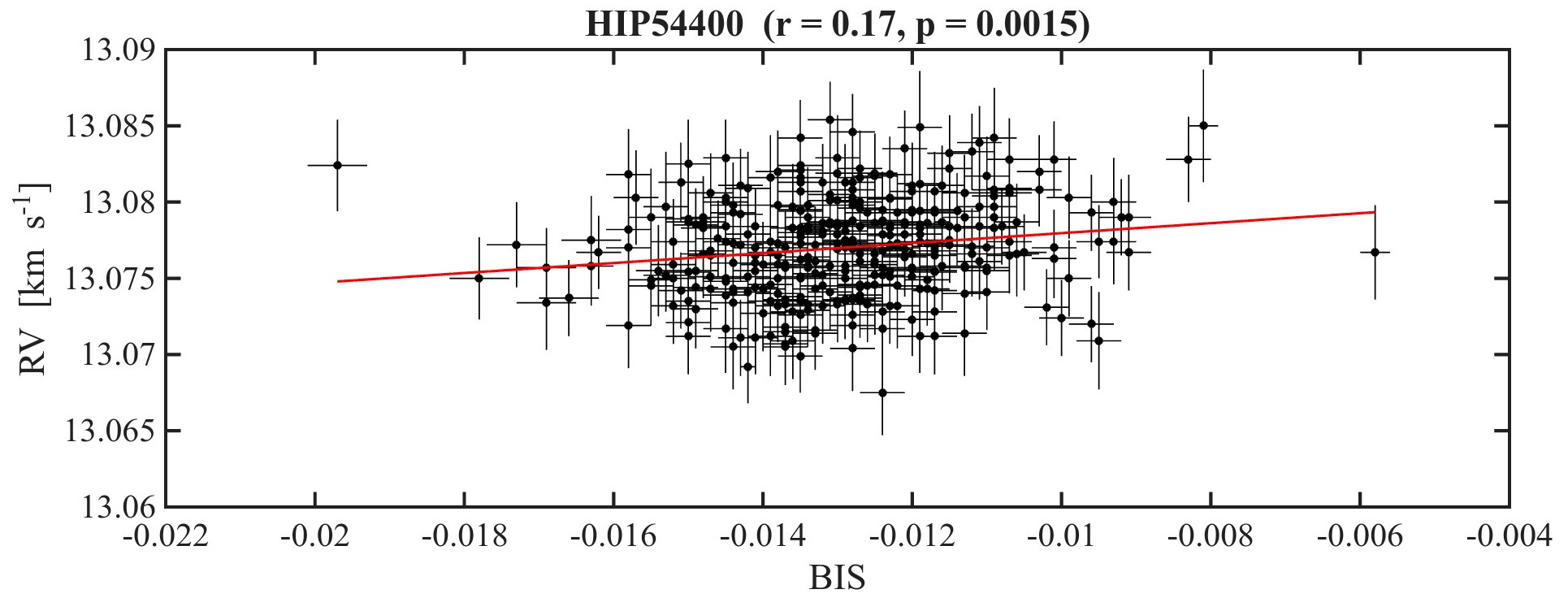}
\end{subfigure}
\begin{subfigure}{.5\textwidth}
\includegraphics[width=\linewidth]{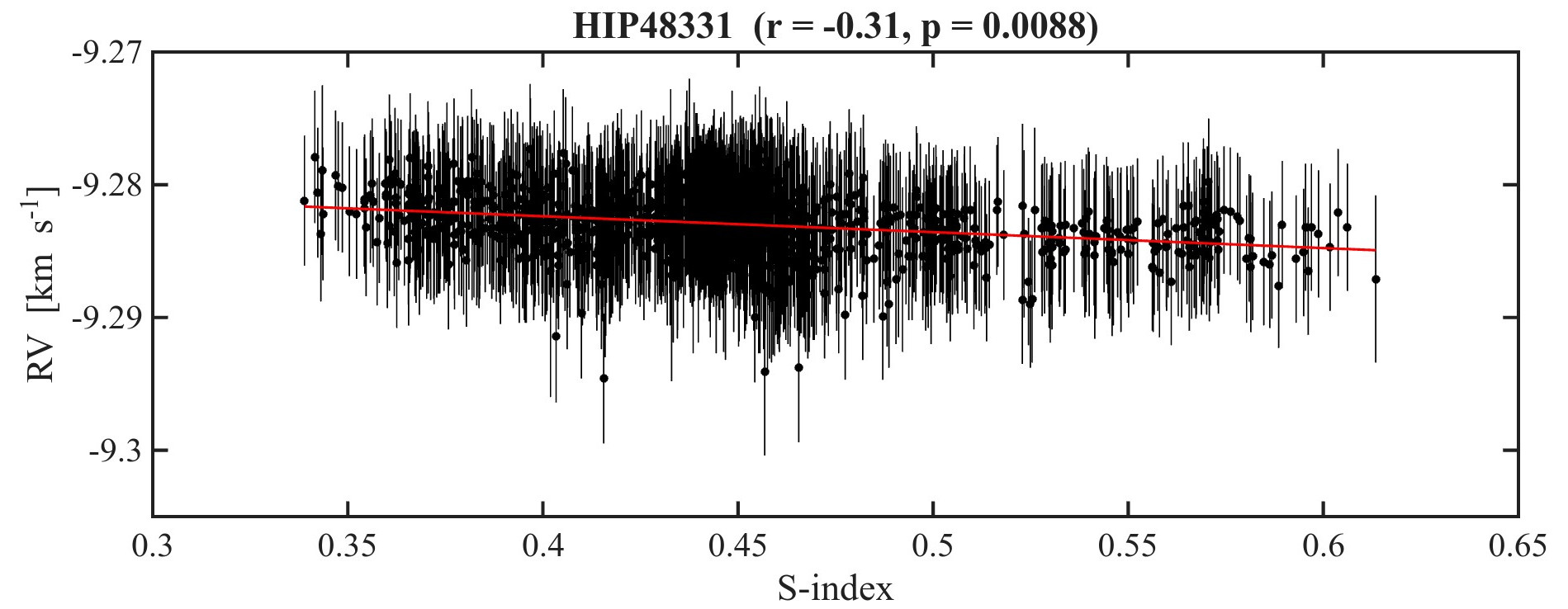}
\end{subfigure}

\caption{RV variations as a function of stellar activity indicators (S-index and BIS). The red lines show linear fits to the data, and the corresponding correlation coefficients and p-values are indicated in the panel titles. Left panels: RV vs BIS for: HIP1499 (top), HIP28460 (middle), and HIP5440 (bottom). Right panels: RV vs S-index for : HIP7240 (top), HIP40693 (middle), and HIP48331 (bottom).}
\label{fig:correlation}
\end{figure}

This appendix presents the correlation between the RVs and the stellar activity indicators (S-index and BIS) in Figure \ref{fig:correlation}.
\section{Rotation periods}
Table \ref{tab:Prot_comp} and Table \ref{tab:Prot_newLit} compare literature rotation periods to activity indicators periods, for stars with and without planetary companion. New possible estimations of rotation periods are listed in Table \ref{tab:Prot_newUnknown}.
\begin{table}
\begin{center}
\tiny
    \caption{\small Rotation periods from literature (Lit $P_{\rm{rot}}$) compared to periods determined from S-index ($P_{\rm{S}}$) and BIS ($P_{\rm{BIS}}$) periodograms for the stars with known planetary companions sample. The number of spectra, $N$, is listed in the last column.}
    \vspace*{5mm}
    \label{tab:Prot_comp}
    \begin{tabular}{rrrrrr}
    \hline
HIP/TYC & Lit $P_{\rm{rot}}$ [d] & Ref.& $P_{\rm{S}}$ [d] & $P_{\rm{BIS}}$ [d]  & N\\
\hline
1499 &$30.2 \pm 3.5$ &\protect\cite{diaz2016harps}& $23.6035 \pm 0.0076$&$21.3744 \pm 0.0042$ & 467\\ 
1692 &$241.3 \pm 1.3$ &\protect\cite{zoghbi2011quantization}& -- & $378.1 \pm 1.0$ & 45\\ 
2350 &$38.8 \pm 4.6$ &\protect\cite{watson2010estimating}  &-- & $32.082 \pm 0.29$ & 39\\
5806 &$13.5 \pm 2.6$ &\protect\cite{lovis2011harps}& $14.1668 \pm 0.0042$ &-- & 131\\
10301 &$38.99 \pm 0.47$& \protect\cite{ahrer2021harps}& $38.1418 \pm 0.0079$& -- & 245\\
14501 &$24.9  \pm 2.5$ & \protect\cite{crepp2014trends}& $64.22 \pm 0.12$ & $69.074 \pm 0.085$ & 70\\
15510 &$33.19 \pm 3.61$ &\protect\cite{pepe2011harps}& $28.79 \pm 0.052$ &$32.09196 \pm 0.00031$ & 6782\\
16085 &$35.2 \pm 4.0$ &\protect\cite{udry2019harps} & $33.4632 \pm 0.0090$& $33.322 \pm 0.052$ & 218\\
17264 &$40 \pm 1$ &\protect\cite{barros2022hd} & $34.6104 \pm 0.0045$ & $78.055 \pm 0.084$ & 40\\
20277 &$44.5 \pm 4.2$ &\protect\cite{watson2010estimating}&$43.9869 \pm 0.0015 $ &$46.591 \pm 0.033$ & 99\\
26394 &$17.3 \pm 1.6$ &\protect\cite{watson2010estimating} &--& $19.922 \pm 0.011$ & 487\\
27080 &$41.4 \pm 4.0$ & \protect\cite{lovis2011harps}& $42.259 \pm 0.062$ & $31.217 \pm 0.033$ & 273\\ 
27435 &$23.6 \pm 3.1$ &\protect\cite{lovis2011harps} & $23.523 \pm 0.021$ &$29.2455 \pm 0.0043$ & 252\\ 
27887 &$47.2 \pm 5.3$ &\protect\cite{lovis2011harps} &$42.695 \pm  0.086$ &--& 449\\ 
29432 &-- &--&  $107.7 \pm 3.9$ & $107.7 \pm 3.3$ & 83\\
30579 &$34.0 \pm 3.7$ &\protect\cite{lovis2011harps} &$35.187 \pm 0.011$&--& 122\\ 
33229 &$39.6 \pm  4.1$& \protect\cite{lovis2011harps} &$ 36.747 \pm 0.010$&--&222\\ 
33719 &$12.30 \pm 0.15$& \protect\cite{ballot2011accurate}& $12.8653 \pm 0.0030$ & --& 109\\
38041 &$26.2 \pm 3.8 $&\protect\cite{lovis2011harps}& $26.46853 \pm 0.00051$ & $18.9126 \pm  0.0085$&130 \\
41529 &$36.5 \pm  4.1$ &\protect\cite{lovis2011harps} & $31.6745 \pm 0.0057$ &--& 153\\
42030 &$20.73 \pm 0.76$ &\protect\cite{watson2010estimating}& $22.172 \pm 0.033$ & $22.1722 \pm 0.0043$ & 58\\
42401& $12.2 \pm 0.2$ & \protect\cite{barragan2022young} & $12.44243 \pm 0.00025$ & $6.016331 \pm 0.000045$ &81\\
47202 &$36.0 \pm 4.4$ &\protect\cite{watson2010estimating}& $32.456 \pm 0.058$ &--&56\\ 
49067 &$26.6 \pm 1.5$ &\protect\cite{hebrard2010harps}& $26.5394 \pm 0.0026$& $26.958 \pm 0.078$& 110\\
52521 &$46.2 \pm 5.3$ &\protect\cite{lovis2011harps}& $44.500 \pm 0.011$& $21.2056 \pm 0.0019$ & 139\\ 
56572 &$39.0 \pm 2.0$ & \protect\cite{naef2007harps}& $32.86 \pm 0.85$ &--& 36\\
58451 &$27.40  \pm 0.10$ &\protect\cite{segransan2011harps}&$29.70 \pm  0.12$ &$33.24 \pm 0.11$ & 88\\
60081 &$32.45 \pm 1.7$ &\protect\cite{watson2010estimating} &$28.69 \pm 0.12$ &$20.5329 \pm 0.0042$ & 105\\ 
62345 &$33.4 \pm 3.5$ &\protect\cite{lovis2011harps}& $25.31 \pm 0.20$&--&40\\
70170 &$37.2 \pm 2.0$ &\protect\cite{suarez2015rotation} & $37.166 \pm 0.010$& $36.99 \pm 0.16$ & 135\\
74273 &$23.0 \pm 2.9$ &\protect\cite{udry2019harps}& $30.254 \pm 0.094$&--&337\\ 
74653 &$40.1 \pm 3.8$ &\protect\cite{lovis2011harps} & $19.1591 \pm 0.0087$ & --& 287\\
82632 &$31.0 \pm 1.0 $ &\protect\cite{barbato2020gaps}& $15.0446 \pm 0.0031$& --&54\\
83541 &$42.6 \pm 4.4$ &\protect\cite{lovis2011harps}  & $37.3925 \pm 0.0089$ &$ 37.4762 \pm 0.0032$&269\\
89354 &$30.0 \pm 5.0$ &\protect\cite{marmier2013coralie} & -- & $15.42459 \pm 0.00057$&54\\
93858 &$28.4 \pm 3.5 $& \protect\cite{lovis2011harps} & -- & $27.9524 \pm 0.0030$&232\\
95740 &$32.0 \pm 1.0$ &\protect\cite{zoghbi2011quantization} &-- &$37.86 \pm 0.19$ & 20\\
98959 &$24.7 \pm 3.2$ & \protect\cite{lovis2011harps} &  $24.685 \pm 0.042$ & --&627\\ 
112190 &$41.8 \pm 5.6$ &\protect\cite{lovis2011harps} & $45.2141 \pm 0.0080$&--&369\\ 
0458-01450-1 &$42.9 \pm  1.0 $ &\protect\cite{zoghbi2011quantization}& $45.22 \pm 0.24$ & --&90 \\
\hline		 
\end{tabular}
\end{center}
\end{table}

\begin{table}
\begin{center}
\tiny
    \caption{\small Same as \ref{tab:Prot_comp} but for the stars with no known planetary companion sample.}
    \vspace*{5mm}
    \label{tab:Prot_newLit}
    \begin{tabular}{rrrrrr}
    \hline
HIP & Lit $P_{\rm{rot}}$ [d] & Ref.& $P_{\rm{S}}$ [d] & $P_{\rm{BIS}}$ [d] & N\\
\hline
1444 & $19.9 \pm 1.4$ & \protect\cite{suarez2015rotation}& $20.174 \pm0.013$ & --&209\\
3170 & $15.8 \pm 2.5$ &\protect\cite{lovis2011harps} &--&$12.776 \pm 0.024$ & 256\\
7585 & $23.0 \pm 1.0$&\protect\cite{bellotti2025bcool} & $28.012 \pm 0.54$ & $26.49  \pm 0.22$ & 80\\
9400 & $29.8 \pm 4.3$& \protect\cite{lovis2011harps}  & $36.081 \pm0.021$ &--& 78\\
29271 & $32.00 \pm 1.0$& \protect\cite{barnes2007ages} &     $  32.065 \pm0.016 $ & --&268\\
42291 & $19.8 \pm 2.8$ &\protect\cite{lovis2011harps}& $28.752 \pm 0.015$ & --&227\\
44811 & $16.5 \pm 2.6$ & \protect\cite{lovis2011harps}& $22.602 \pm 0.029$ & --&71\\
64408 &$34.1  \pm 3.5$ &\protect\cite{lovis2011harps} & $32.0574 \pm 0.0020$& --&736\\
69972 & $42.0 \pm 5.9$ &\protect\cite{lovis2011harps}  &$40.488 \pm 0.024$&$ 20.556\pm 0.036$&91\\
74389 & $23.2 \pm 3.2$ & \protect\cite{lovis2011harps}& $42.648 \pm 0.031$ & --&44\\
75253 & $45.4 \pm 5.6$ &\protect\cite{lovis2011harps}  &$37.2688 \pm0.0073 $ & $55.13 \pm 0.42$&72\\
80683 & $28.9 \pm 5.6$ & \protect\cite{lovis2011harps} &$28.890 \pm0.014$ & $36.993 \pm 0.083$&73\\
83990 & $45.8 \pm 5.0$ & \protect\cite{lovis2011harps} & $50.3280 \pm0.0078$ &--&755\\ 
87369 & $22.4 \pm 3.8$ & \protect\cite{lovis2011harps} & $26.62 \pm0.21$&--&17\\ 
91700 & $28.3 \pm  3.9$ & \protect\cite{lovis2011harps}& $47.26 \pm0.45$ & $23.526 \pm 0.045$&43\\
100233 & $17.5 \pm 2.7$ &\protect\cite{lovis2011harps}  &$22.77\pm  0.15$ &--&65\\
101997 & $33.1 \pm 3.7$ & \protect\cite{lovis2011harps} &$24.47 \pm 0.28$ & $28.672 \pm 0.015$&64\\ 
103458 & $18.1 \pm 2.7$ &\protect\cite{lovis2011harps}& -- & $24.34891 \pm 0.00030$ &1166\\
117320 & $27.9 \pm 3.2$ &\protect\cite{lovis2011harps} &  $35.6311 \pm0.0012$ & --&252\\
\hline		 
\end{tabular}
\end{center}
\end{table}

\begin{table}
\begin{center}
\tiny
    \caption{\small Possible new estimations of rotation periods from S-index ($P_{\rm{S}}$), BIS ($P_{\rm{BIS}}$) and Tess ($P_{\rm{TESS}}$) periodograms. The number of spectra, $N$, is listed in the last column.}
    \vspace*{5mm}
    \label{tab:Prot_newUnknown}
    \begin{tabular}{rrrrr}
    \hline
HIP & $P_{\rm{S}}$ [d] & $P_{\rm{BIS}}$ [d] & $P_{\rm{TESS}}$ [d]& N\\
\hline
57 & $16.9587 \pm 0.0024$ & --& --& 37\\
80 & $10.7828 \pm 0.0433$& $5.04757 \pm 0.0010$ &--& 121\\
569 & $45.02 \pm 0.29$& $42.12 \pm 0.10$&--& 41 \\
5938 & $3.6897 \pm 0.0015$& $3.679758 \pm 0.000097$&$7.2 \pm 1.6$& 30\\
8119 & $20.966 \pm 0.059$ & --&$17.938 \pm 0.021$&12\\
12109 & --& $8.0870 \pm 0.0023$ & $9.68 \pm 0.11$& 71\\
12110 & $4.5836 \pm 0.0061$ & $10.543 \pm 0.027$ &  $11.97 \pm 0.19$& 18\\
14587 & -- & $7.2798 \pm 0.0025$ & $ 8.592 \pm 0.055$ & 26\\
30260 & -- & $14.18 \pm 0.18$&--& 32\\
30979 & -- & $14.629 \pm 0.053$&--&19\\
31712 & $28.33 \pm 0.11$ & $29.277 \pm 0.028$&--& 60\\
36941 & $5.5535 \pm 0.0023$ & $5.5540 \pm 0.0017$&--&68\\
37447 &$2.1493 \pm 0.0022$ & $2.1683 \pm 0.0046$&--&41 \\
43797 & $2.5050 \pm 0.0058$ & $2.3211210 \pm 0.00014$&--&44\\
47681 & $16.43 \pm 0.15$ &--&  $8.1 \pm 2.4$&25\\
50534 & -- & $7.978 \pm 0.013$ & $8.1 \pm 2.0$&50\\
54102 & $10.904 \pm 0.011$ &--& $8.38 \pm 0.13$ &29\\
66547 & -- & $3.66462 \pm 0.00058$ &  $6.42 \pm 0.45$ &24\\
79715 & -- & $53.20 \pm 0.15$ &--& 50\\
81533 & $106.2 \pm 1.8$ & $97.6 \pm 4.9$ &--& 20\\
81819 &$ 45.486 \pm 0.026$  & --&--&56\\
91805 & $16.016 \pm 0.083$ & $14.956 \pm 0.027$&--&103\\
96536 & $7.35 \pm 0.11$ & $7.059 \pm 0.010$&--&39\\
99174 & $18.129 \pm 0.044$ & --&--&25\\
105184 & $23.0430 \pm 0.0013$ & $26.894020 \pm 0.000013$&--&175 \\
108241 & $17.27 \pm 0.13$ & $17.85 \pm 0.48$&--&30\\
112491 & $8.79333 \pm 0.00036$ & $8.914 \pm 0.0024$&--&103\\
\hline		 
\end{tabular}
\end{center}
\end{table}
\section{Magnetic cycle periods}
Table \ref{tab:Pmag_compLit} and Table \ref{tab:Pmag_newLit} compare magnetic cycle periods to activity indicators periods, for stars with and without planetary companion. New possible estimations of magnetic cycle periods are listed in Table \ref{tab:Pmag_compUnknown} and Table \ref{tab:Pmag_newUnknown}.
\begin{table}
\begin{center}
\tiny
    \caption{\small Magnetic cycle periods from literature ($P_{\rm{mag}}$) compared to periods determined from S-index ($P_{\rm{S}}$) and BIS ($P_{\rm{BIS}}$) periodograms for the stars with known planetary companions sample. The number of spectra, $N$, is listed in the last column.}
    \vspace*{5mm}
    \label{tab:Pmag_compLit}
    \begin{tabular}{rrrrrr}
    \hline
HIP & Lit $P_{\rm{mag}}$ [d] & Ref. & $P_{\rm{S}}$ [d] & $P_{\rm{BIS}}$ [d]& N \\
\hline
7599 & $1498 \pm 146$& \protect\cite{mascareno2016magnetic}& $4919 \pm 365$& --&328\\
10301 & --&--& $1276.8 \pm 3.3$& --&245\\
11433 & $1111 \pm 396$& \protect\cite{lovis2011harps}& $2781 \pm 23$& -- &116\\
14530 & $2812 \pm 183$& \protect\cite{boro2018chromospheric}& $3441 \pm 28$& $2833 \pm 25$&198\\
27887 & $3804 \pm 806$& \protect\cite{lovis2011harps}& $4571 \pm 16$& $3985 \pm 19$&449\\
30503 & $1790\pm 110$& \protect\cite{boro2018chromospheric}& $4194 \pm 611$& --&324 \\
33229 & $2481 \pm 322$& \protect\cite{lovis2011harps}& $4506 \pm 57$& --&222\\
38041 & $4712 \pm 1790$& \protect\cite{mascareno2016magnetic}& --& $1101.1 \pm 1.5$ &130\\
41529 & $3222 \pm 565$& \protect\cite{lovis2011harps}& $3793 \pm 32$& $3805 \pm 58$ &153\\
48331 & $3793 \pm 806$& \protect\cite{lovis2011harps}& $3850.5 \pm 1.1$ & $2225.62 \pm 0.90$&1191 \\
48331 & --&--& $1188.66 \pm 0.30$ & $1939.8 \pm 1.2$ & 1191\\
52521 & $3453  \pm 806$& \protect\cite{lovis2011harps}& $4262 \pm 22$& $3656 \pm 17$ &139\\
70170 & $2155 \pm 219$& \protect\cite{mascareno2016magnetic}& $5385 \pm 399$& --&135\\
72119 & $1419  \pm 436$& \protect\cite{lovis2011harps}& --& $2351 \pm 10$ & 137\\
93858 & $2016 \pm 355$& \protect\cite{lovis2011harps}& $1979.4 \pm 1.2$& $3225.2 \pm 8.2$ &232\\
106006 & $987 \pm 73$& \protect\cite{lovis2011harps}& --& $3539 \pm 365$ & 112\\
\hline		 
\end{tabular}
\end{center}
\end{table}

\begin{table}
\begin{center}
\tiny
    \caption{\small Possible new magnetic cycle periods detected from S-index ($P_{\rm{S}}$) and BIS ($P_{\rm{BIS}}$) periodograms for stars with known companion. The number of spectra, $N$, is listed in the last column.}
    \vspace*{5mm}
    \label{tab:Pmag_compUnknown}
    \begin{tabular}{rrrr}
    \hline
HIP &  $P_{\rm{S}}$ [d] & $P_{\rm{BIS}}$ [d] & N\\
\hline
801 & $1568.5 \pm 7.5$& --&107\\
3497 &  $2265 \pm 451$& --&362\\
12653 & $5021 \pm 371$& -- &2118\\
17264 & $2776 \pm 60$& --&40\\
26013 & $2496 \pm 43$& --&172\\
32970 & --& $1630 \pm 803$&39\\
39417 & $2441 \pm 460$& -- &28\\
40693 & $4460 \pm 13$& $3937.0 \pm 3.5$&741\\
40693 & $1323.3 \pm 5.3$& --&741\\
40952 & $3937 \pm 770$&  $2286 \pm 402$ &39\\
44291 & $3238 \pm 493$ & --&160\\
47007 & $1708.8 \pm 5.7$& $1601.8 \pm 1.8$&255 \\
47202 & $2177 \pm 226$& -- &56\\
49067 & $4513 \pm 85$& -- &110\\
57931 & $4401 \pm 43$& -- &64\\ 
74273 & $1574.3 \pm 3.7$& --&337\\ 
74500 & --& $4100 \pm 40$&131 \\ 
90979 & $2470 \pm  89$& -- &44\\
93540 & $2303 \pm 15$& -- &169\\
98959 & --& $4601 \pm 365$&627\\
108375 & $2693 \pm 37$& --&126 \\ 
112441 & $1209.5 \pm 5.0$& --&114\\ 
\hline		 
\end{tabular}
\end{center}
\end{table}

\begin{table}
\begin{center}
\tiny
    \caption{\small Same as Table. \ref{tab:Pmag_compLit} but for stars with no known companion.}
    \vspace*{5mm}
    \label{tab:Pmag_newLit}
    \begin{tabular}{rrrrrr}
    \hline
HIP & Lit $P_{\rm{mag}}$ [d] &Ref. & $P_{\rm{S}}$ [d] & $P_{\rm{BIS}}$ [d]& N \\
\hline
1382 & $1205 \pm 73$ & \protect\cite{mascareno2016magnetic} & $1202 \pm 781$ & --&20\\
1599 & $1018 \pm 51$ & \protect\cite{lovis2011harps}& $1062 \pm 32$ & --&3834 \\
1599 & -- &-- & $5441 \pm 412$ & -- &3834\\
55210 & $2861 \pm 1354$ & \protect\cite{lovis2011harps} & $3849 \pm 345$& --&90 \\
69972 & $1146 \pm 982$ & \protect\cite{lovis2011harps} & $2862 \pm 20$ & $2488 \pm 12$&91 \\
80683 & $2416 \pm 936$& \protect\cite{lovis2011harps} &$ 2988 \pm 32$ & --&73\\
82588 & $4372  \pm 37$& \protect\cite{boro2018chromospheric} & $1104 \pm 20$ & --&116\\
83990 & $2702 \pm 412$& \protect\cite{lovis2011harps} &  $2594.8 \pm 2.2$ & -- &755\\ 
85425 & $1424 \pm 183$& \protect\cite{boro2018chromospheric} & $1750 \pm 43$ & -- &68\\

\hline		 
\end{tabular}
\end{center}
\end{table}

\begin{table}
\begin{center}
\tiny
    \caption{\small Same as Table. \ref{tab:Pmag_compUnknown} but for stars with no known companion.}
    \vspace*{5mm}
    \label{tab:Pmag_newUnknown}
    \begin{tabular}{rrrr|rrrr}
    \hline
HIP & $P_{\rm{S}}$ [d] & $P_{\rm{BIS}}$ [d]&N &HIP & $P_{\rm{S}}$ [d] & $P_{\rm{BIS}}$ [d]&N\\
\hline
57 & -- & $1307 \pm 207$&37 & 63157 & $4506 \pm 39$ & --&41 \\ 
436 & -- & $3385 \pm 14$ & 80&64550 & $1310 \pm 29$ & --&144\\ 
436 & -- & $1695.0 \pm 7.2$& 80&74389 & $1662 \pm 180$ & --&44\\
948 & $4938 \pm 238$ & --&26&75253 & $2884 \pm 25$ & --&72 \\
2747 & -- & $1020.2 \pm 8.1$&59 &76200 & $2101 \pm 104$ & --&23 \\
8361 & $3597.7 \pm 3.0$ & -- &42&78170 & $3492 \pm 67$  & --&51 \\
8674 &$ 2928.5 \pm 4.8$ & -- &75&78955 &$ 3661 \pm 221$ & $4359 \pm 2611$&171 \\ 
8830 & $3551 \pm 72$ & -- &41&85042 & -- & $3261 \pm 425$&201\\ 
11915 & $2807 \pm 138$ & $3849 \pm 245$&96 &86765 & $2708 \pm 100$ & $2583 \pm 45$&154 \\ 
12109 & $4623 \pm 238$ & -- &71&89211 & -- & $1180 \pm 22$&25 \\
12350 & $2544 \pm 66$ & -- &46&90656 & $3384 \pm 127$ & $3256 \pm 193$&51 \\ 
14614 & $3370 \pm 1231$ & -- &32&96861 & $3303 \pm 19$ & -- &75\\ 
14684 & $1136 \pm 27$ & -- &39&98106 & $4138 \pm 837$ & --&35\\ 
19925 &  $1362 \pm 88$ & -- &45&98589 & $1364 \pm 116$ & -- &59\\
22504 & $1194 \pm 191$ & --&22&100223 & $2022 \pm 68$ & $1872 \pm 230$ &46\\ 
26973 & $1038.4 \pm 9.5$ & --&46 &101997 & $3724 \pm 34$ & -- &64\\ 
35902 & $3422 \pm 113$ & --&25 &102203 & $2148 \pm 548$ & -- &19\\ 
36512 & $2192 \pm 936$ & --&43&103458 & -- & $1981.9 \pm 3$&1166 \\ 
44997 & $2875 \pm 229$ & --&33&103572 & $973 \pm 11$ & --&114 \\ 
45533 & $2624 \pm 10$ & --&102&103682 & $1663 \pm 73$ & --&76 \\
48594 & $2623 \pm 126$ & --&25&103768 & $3214 \pm 154$ & -- &25\\
50075 & $1085 \pm 27$ & -- &52&104045 & -- & $3977 \pm 475$&47\\
52369 & $1659 \pm 38$ & -- &90&108468 & -- & $2850  \pm 1242$&67 \\
53087 & $3217 \pm 154$ & -- &54&109166 & $2336 \pm 112$ & -- &26\\
57507 & $8562.7 \pm 2.6$  & --&136&109821 & -- & $2064.3 \pm 7.5$ &250\\
58289 & $3875 \pm 186$ & -- &28&115445 & $1972 \pm 79$ & -- &35\\ 
62107 & $1536 \pm 59$  & --&26&116763 & $1884 \pm 52$ & -- &60\\
\hline		 
\end{tabular}
\end{center}
\end{table}

\clearpage
\bibliographystyle{raa}
\bibliography{bibtex}



\label{lastpage}

\end{document}